# A predictive scaling framework for hierarchical living systems: From biomacromolecule to the Anthropogenic Earth system

Jie Chang[a*], Ying Ge[a], Xiaoyu Hou[a], Zhaoping Wu[b], Yong Min[c], Feng Mao[d], Guofu Yang[e]

a College of Life Sciences, Zhejiang University, Hangzhou, 310058, China

b School of Public Administration, Zhejiang University of Technology, Hangzhou, 310023, China

c Computational Communication Research Center, Beijing Normal University, Zhuhai, 519087, China

d School for Cross-Faculty Studies, University of Warwick, Coventry, CV4 7AL, the United Kingdom

e School of Art and Archaeology, Hangzhou City University, Hangzhou, 310016, China

*Author for correspondence. E-mail: jchang@zju.edu.cn

**Abstract**

Living systems, from biomacromolecules to the biosphere, are organized hierarchically with quantitative traits (such as size, mass and lifespan) spanning many orders of magnitude. However, qualitative traits such as structure differ between adjacent levels, and similarities among non-adjacent levels have been recognized only case by case. Yet no framework integrates quantitative and qualitative traits across the hierarchy of living systems. Here, we propose HELIX, a model that couples exponential scaling across levels with trait-based classification. Clustered by 16 traits, the 14 known levels are classified into three groups: Autonomous, Component, and Repeat-autonomous living systems. Levels within each group are self-similar, and the three groups alternate periodically in a helical belt and the quantitative traits follow scaling law. HELIX integrates existing principles from multiple disciplines to form a unified theory. HELIX further predicts three newly emerged levels, illustrating that it can guide cross-level and interdisciplinary research on living systems.

**Keywords** living system group, periodic, quantitative trait, qualitative trait

## 1 Introduction

A hierarchical system is a multi-level, nested structure composed of a series of systems ordered from small to large (Odum, 1971; O'Neill et al., 1989; Odum and Barrett, 2005; Pavé, 2006; Jagers op Akkerhuis, 2008; Cottam and Vounckx, 2021). The levels of organization range from quantum, atoms to molecules in physics, through living cells, organisms, ecosystems to biosphere in biology and ecology (Odum, 1971; Weiss, 1969; Chang and Ge, 2001, 2005; Korn, 2002; West, 2017), to the Solar System, stars and galaxies in cosmology (Koestler, 1978; Haber, 1994). As a fractal system, a hierarchical system should allow traits to scale up and down across levels, following the rule of self-similarity (Mandelbrot, 1982; Chang et al., 1996; West et al., 1999). For example, metabolic rate scales with mass as a power law within each major group of living systems, from prokaryotes and protists to multicellular organisms and eukarcities (DeLong et al., 2010; Wu et al., 2025). Yet, scaling the qualitative features (structure and process) of living systems between adjacent levels has not yet been achieved. For example, the morphology, structure, development, metabolism and homeostasis of an organism differ from those of a population at a higher level or an organ at a lower level (Chang and Ge, 2001; Reuter et al., 2010; Chang et al., 2021).

Fortunately, similarities in certain qualitative features have been found between some non-adjacent levels of living and physical systems. In physics, the famous 'Atomic Planet Model' draws an analogy between the atomic structure and planetary systems (Rutherford, 2012; Dylla, 2020). The similarities among several types of systems, such as multistage river networks, crystals, plant vascular networks and plant root systems, follow fractal patterns (Mandelbrot, 1982; Chang et al., 1996; Chang and Ge, 2001, 2005; West, 2017). The Logistic Model of population in ecology has also been applied to describe the growth of cell populations, tumours and cell cultures (Kozusko and Bourdeau, 2007; Warne et al., 2017). The GAIA hypothesis, which reveals the steady-state regulation of the Earth's energy and chemical conditions (Lovelock and Margulis, 1974), has been likened to the homeostasis of mammals and birds (Chang and Ge, 2005, 2010). The super-cell city hypothesis (Chang et al., 2021) has found many shared traits between eukaryotic cells and modern cities. Because systems with similar traits occur between non-adjacent levels across the biological spectrum, we hypothesize that hierarchical models should have a more comprehensive and reasonable form.

Another observed critical phenomenon is periodicity, both intra-level or inter-levels. At the atomic level, the most famous example is Mendeleev's periodic table, in which elements in the same group share similar chemical properties, while atomic radius changes systematically across each period and increases from one period to the next, as atomic mass rises steadily (Pyykkö, 2019). At the molecular level, individual organic molecules in solution spontaneously form supramolecular complexes periodically (Dhar, 2007; Babaev, 2019). Within hierarchical systems, periodicity spans multiple levels, from superstrings (the minimum physical level) to multicellular organisms (Jagers op Akkerhuis, 2008). There are structural similarities between non-adjacent levels of living systems, and this suggests that hierarchical living systems exhibit periodicity (Chang and Ge, 2001, 2005). However, present models fall short in measuring cross-level characteristics, such as the increase of length and mass from one period to the next.

The number of level of living systems in the hierarchical spectrum has been increasing along with evolution. In the starting of the "origin of life" approximately 3.6 billion years ago, there is only prokaryote cell and the components. The second great emergence of life in evolution is eukaryotic cell, and then many levels of living systems continue increasing. Margulis (2010) hypothesized the endosymbiosis of living systems, identifying three fusions: the emergence of prokaryotic cells, eukaryotic cells, and multicellular organisms. Chang et al. (2021) hypothesized that the eukarcity (a coupled urban and rural system emerging after the 3$^{rd}$ Industrial Revolution) represents the 4$^{th}$ fusion. A question is, are there other fusions among living systems? Besides these fusions, many things emerged in the Anthropocene, such as human cultural information, termed meme (Dawkins, 1976), or cene (Chang and Ge, 2005), has been regulating the Earth system. Human-made materials have surpassed present biomass of the Earth around 2020 (Elhacham et al., 2020). The energy harnessed by humans has dominated the social-ecological systems that provide most ecosystem services for humanity. Yet, no quantitative models exist for the evolution of the hierarchical living system spectrum in general.

In this study, we first build a one-dimension model (1D) by regressions and find the exponential equations of the scale traits (length, volume, mass and time) of the 13 known levels of living systems. For 16 traits of structural, process and functional (e.g., outer membrane, control center and metabolic scaling exponent), we build a two-dimension model (2D) using cluster analysis to rearrange the 14 living systems into three groups. We then integrate the two models into a three-dimensional model (3D), with the 1D scale axis perpendicular to the 2D cluster plane. The levels thus spiral upward through the three groups, forming a helical curve, which we name HELIX. The 1D model reveals gaps in the scale sequence, and HELIX predicts that three unknown levels of living systems should occupy these positions. Next, we identify real-world candidate systems at the predicted positions, test whether their observed traits fall within the 95% prediction intervals of the 1D and 3D models fitted to the 14 known levels, and re-cluster them with the 16 traits in the 2D model to test whether their trait profiles match the predicted group. HELIX also accommodates several existing models of hierarchical living systems. Together, these predictions and syntheses establish HELIX as a theoretical framework, which supports the endogeneity of hierarchical living systems. Furthermore, HELIX identifies three newly emerged living systems.

## 2 Material and methods

### 2.1 Data collection and calculation of the quantitative traits

For each of the 14 known levels, we compiled four quantitative traits: length, volume, wet mass and time (lifespan or process). Values were taken from published measurements where available. Otherwise, they were derived from related quantities using standard physical relations; for example, subunit volume was estimated from molar mass, and approximately spherical systems were treated as spheres to obtain length. Data sources, equations and intermediate values are given in Supplementary

Text S1.

Energy metabolism data for prokaryotes, unicellular eukaryotes, multicellular organisms and eukarcities were taken from Wu et al. (2025); details are given in Supplementary Text S2.

**2.2 Selection and standardization of the traits**

For the 14 known levels, we selected 16 traits describing structure (including morphology), process and function: 3 continuous traits (e.g., morphology diversity) and 13 categorical traits (e.g., presence of an outer membrane) (Table S2). Trait definitions followed the Phenotype And Trait Ontology (PATO; Gkoutos et al., 2018) and the Systems Biology Ontology (SBO; Courtot et al., 2011). Cross-referenced with established domain ontologies for each level, from ChEBI and GO for biomacromolecules to ENVO for ecosystems are given in Table S3. For organaras, eukarcities and the biosphere, which lack existing ontologies, we developed new ones following the construction principles of cell and organelle ontologies (https://github.com/smanalysis/Ontology). Continuous traits were $\log_{10}$-transformed where appropriate, and categorical traits were scored on binary or ordinal scales (Tables S4 and S5). Details are given in Supplementary Text S3.

**2.3 Statistical analysis and modeling**

Regressions were performed in Origin (version 2024, OriginLab Corporation, Northampton, MA, USA). Goodness of fit was evaluated by the coefficient of determination ($R^2$), and significance was set at $p < 0.05$.

For the 1D model, $\log_{10}$-transformed traits were regressed on level rank by ordinary least squares; allometric relationships between traits were fitted as power laws, both as a single model across all levels and as piecewise models with breakpoints set a priori at two major transitions: from unicellular to multicellular systems, and from ecosystems to organaras.

For the 2D model, we used a hierarchical cluster analysis with the complete linkage method based on Gower distance (Maechler et al., 2019) to identify similarity among living systems. Cluster analysis was performed in R (version 4.4.2). The number of clusters was determined by cutting the dendrogram at height 0.6.

The 3D model combined the 1D scale axis with the 2D cluster plane. Within each of the three groups, $\log_{10}$-transformed traits were further regressed on period to describe the scaling of recurring system types.

For each newly emerged level, 95% prediction intervals were calculated from both the 1D regressions on level rank and the within-group 3D regressions, fitted to the known levels only.

## 3 Building a HELIX: scaling across all levels and recurrence across groups

### 3.1 To build a quantitative one-dimensional hierarchical model for 14 levels of living systems

The hierarchy of living systems is multi-level, nested, self-similar, diverse, complex and identifiable, with levels ordered from small to large. Living systems at different levels have specific spatial and temporal scales (see Supplementary Text S4 for details). According to the criteria, only 14 levels of living systems (Fig. 1a) meet these criteria. The 14 levels are not evenly spaced along the four trait axes (Fig. 1a). There are gaps between biomacromolecules and cells (all four traits), prokaryotic and eukaryotic cells (mass), unicellular eukaryotes and tissues (length, volume and mass), multicellular organisms and organara (length and volume), and eukarcity and biosphere (length, volume and time). The first two gaps coincide with major evolutionary transitions: the origin of cells and of multicellularity. The gap above the eukarcity may instead contain levels that have emerged only recently. Because this gap implies that intervening levels are missing, the biosphere's rank in the known sequence is underestimated; we therefore excluded it from the 1D and 3D fits.

The length of biomacromolecules is ~$10^{-9}$ m on average and the diameter of the biosphere area is ~$10^{7}$ m on ground surface (Table S6). The 14 levels of living system span lengths from $10^{-9}$ m to $10^{7}$ m (Fig. 1b). The volume of entire system spectrum spans from ~$10^{-25}$ to ~$10^{19}$ $m^3$, 44 orders of magnitudes (Fig. 1c). The mass, which is the wet weight, of entire system spectrum spans from ~$10^{-20}$ to $10^{18}$ g, ranging 38 orders of magnitudes (Fig. 1d). The mean active duration (time) of living systems in the entire spectrum ranges from $10^{-4}$ to ~$10^{11}$ s, 15 orders of magnitudes (Fig. 1e).

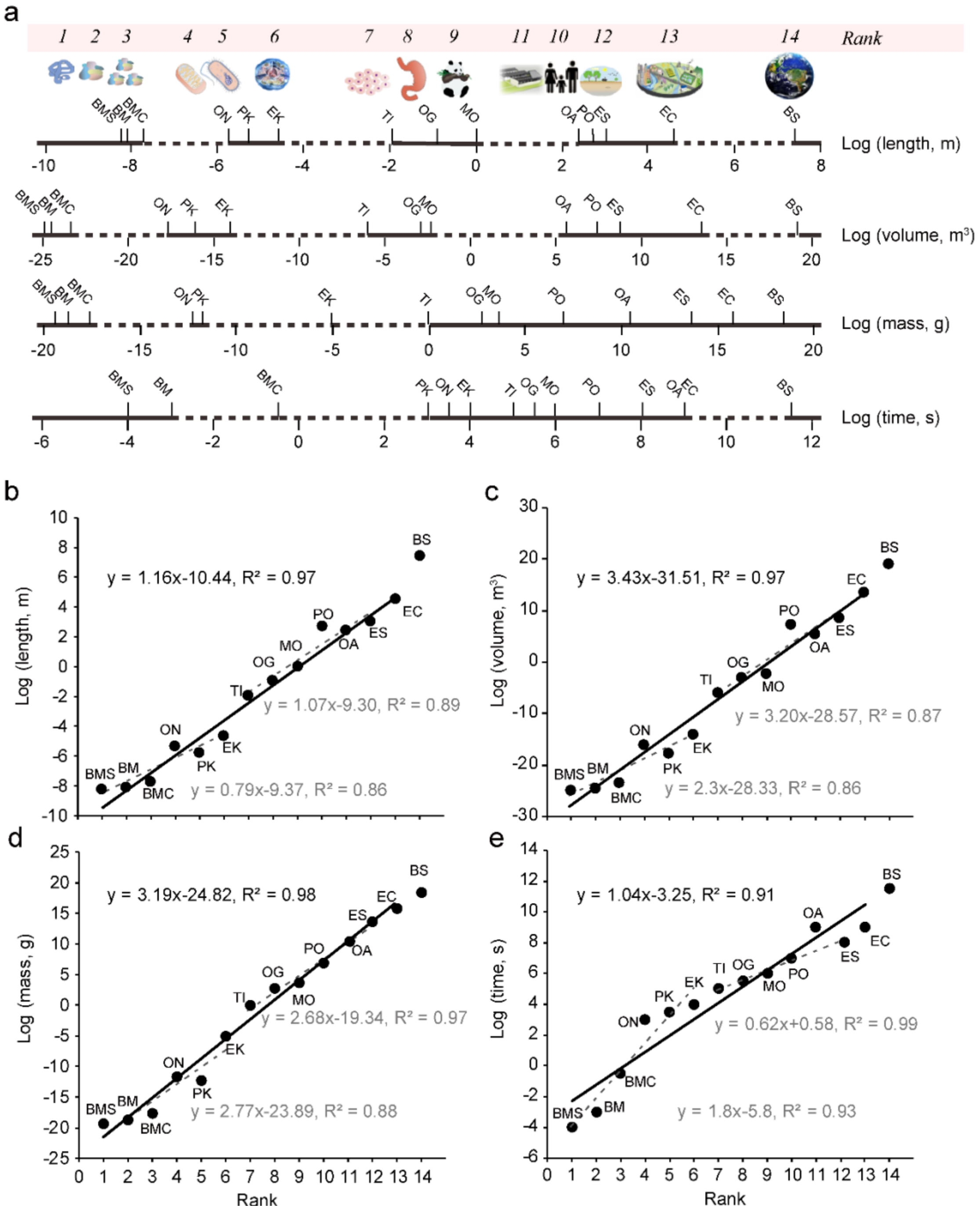

**Fig. 1 Quantitative traits of the 14 levels living systems and their exponential relationships in hierarchy.** (a) Position of the 14 levels on length, mass, volume and time, black dashed segments mark gaps between adjacent levels, gap defined as intervals between adjacent levels exceeding 1.5 folds the mean adjacent interval for that trait (Table S6), numbers in the top row give the hierarchical rank, (b-e) relationships between the rank and length (b), volume (c), mass (d), and time (e), solid lines indicate models fitted to levels 1−13 (n = 13), and grey dashed lines show the piecewise models. Abbreviations and rank: 1 BMS, biomacromolecular subunit, 2 BM,

biomacromolecule, 3 BMC, biomacromolecular complex, 4 PK, prokaryote cell, 5 ON, organelle, 6 EK, eukaryotic cell, 7 TI, tissue, 8 OG, organ, 9 MO, multicellular organisms, 10 PO, population of MO, 11 OA, organara, 12 ES, ecosystem, 13 EC, eukarcity, 14 BS, biosphere. All fitted models are significant ($p < 0.05$).

Regression analysis showed that all the relationships between rank and the size ($S$), including length, mass, volume and time of the living systems are all fit the exponential function, and we define it to be a 1D model for hierarchical spectrum of living systems,

$$S_i = S_0 \cdot \lambda^i \ or \ log\ (S_i) = log\ (S_0) + \ i \cdot log\ \lambda \quad (1)$$

where, $S_i$ is size (length, volume, mass, or active duration) of the living system at hierarchical rank $i$ ($i = 1$ for biomacromolecular subunit), $\lambda$ is the finite rate of increase, representing the multiplicative factor by which size increases per level (where $log\ \lambda$ corresponds to the slope of the regression in Fig. 1).

The 1D model (Eq. 1) shows that as the size increases accelerate along with the level increasing of the living systems. The $\lambda$ of the length between each level is $10^{1.2}$ (Fig. 1b). The $\lambda$ of the volume between levels is $10^{3.4}$ (Fig. 1c). The $\lambda$ of mass between levels is $10^{3.2}$ and the $\lambda$ of time between the levels is $10^{1.0}$ (Fig. 1d, e). However, each of the four quantitative traits can be divided by two and more models with the gaps, and the allometric relationships between traits show the same breaks (Fig. S1; Supplementary Text S5). There are two major segments with a break between the unicellular and multicellular living systems. This is supported by the shift in metabolic scaling exponents across the major evolutionary transitions from unicellular to multicellular life (DeLong et al., 2010), suggesting that the emergence of multicellularity changed the scaling rules of living systems. In addition, the biosphere deviates from the scaling relationships followed by the other 13 levels, because the gap between the eukarcity and the biosphere.

Until now, we have built a one-dimensional (1D) hierarchical living system model, which contains the level scaling models for quantitative traits. The 1D model reveals that length, volume, mass and time scale exponentially with hierarchical level from the biomacromolecular subunit to the eukarcity. The 1D model uncovers the first-class structure of hierarchical living systems, similar to the model for the first-class structure of protein molecules (Dill and MacCallum, 2012). However, the 1D model does not cover qualitative traits such as the shape and number of components of living systems, and these traits cannot be extrapolated between adjacent levels.

### 3.2 To build a two-dimensional group model for the 14 levels of living systems

To find the similarities between the living systems of any levels, we conducted a cluster analysis. We used 16 traits selected based on ontologies (Table S2-S4) of the 14 levels of living systems. Three clusters were identified (Fig. 2a). Cluster 1 contains the biomacromolecular subunit, organelle, organ, ecosystem, organara and biosphere; cluster 2 contains biomacromolecule, prokaryote cell, eukaryotic cell, multicellular organism, and eukarcity; cluster 3 contains biomacromolecular complex, tissue and population (Fig. 2b).

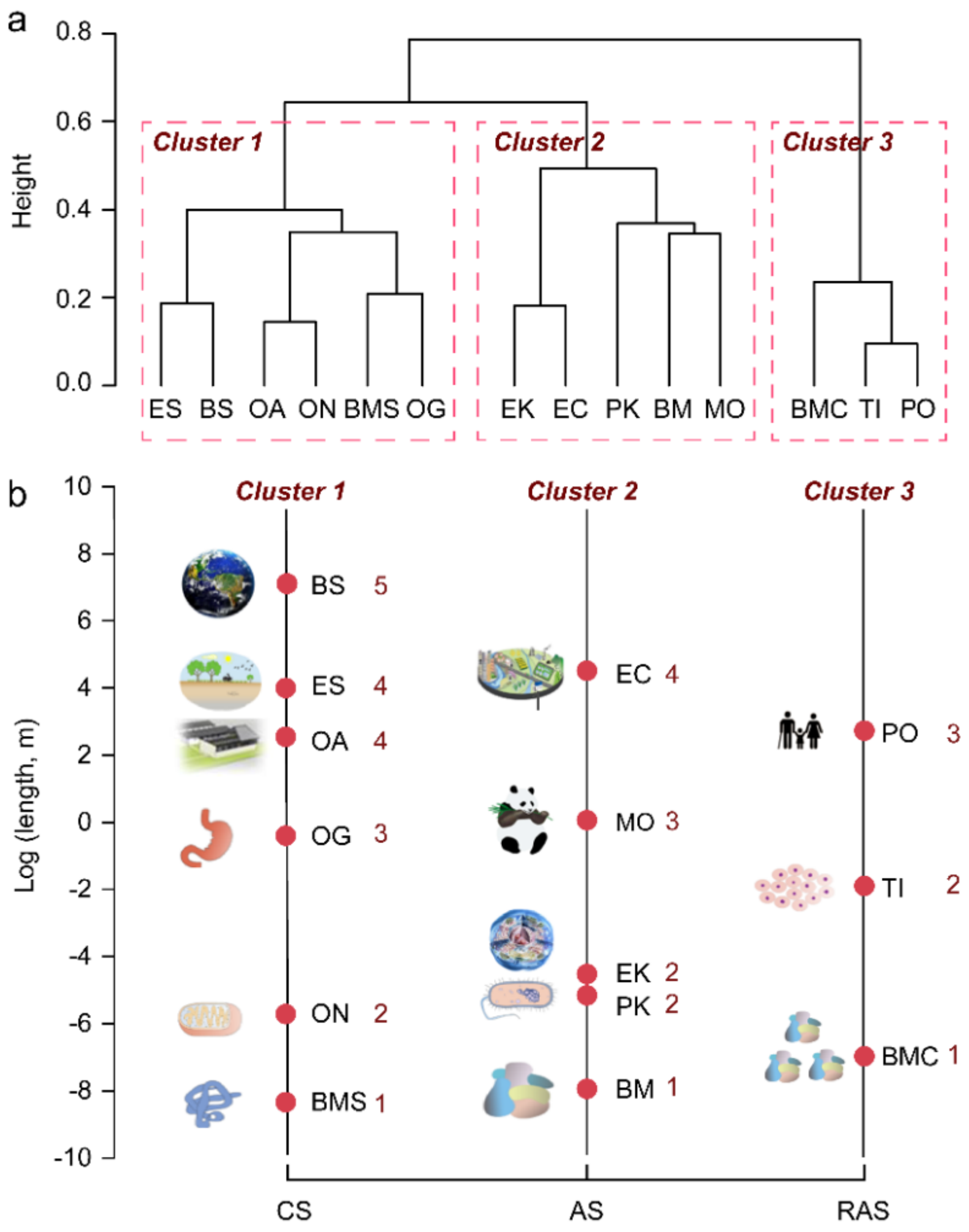

**Fig. 2 Cluster analysis of 14 levels of living systems based on 16 structural, process and functional traits.** (a) Clustering dendrogram, the dashed line marks the cut height of 0.6, which yields three clusters: cluster 1, cluster 2, and cluster 3. (b) Levels arranged by group and plotted against $\log_{10}$ length (m), dark red numbers to the right of system names denote the hierarchical rank within each group. Abbreviations see Fig. 1.

**Similarities of the living systems within the same group.** The living systems within each group share many qualitative traits (*Tql*),

$$Tql_{k,g,j} \approx Tql_{g,j} \tag{2}$$

Where $Tql_{k,g,j}$ is qualitative trait $j$ at level $k$ in a group $g$, $Tql_{g,j}$ is the shared trait $j$ in group $g$.

Members of each group share key qualitative traits, as expressed by Eq. (2) (Tables S5, S7-S9). In cluster 1, for example, organelles (e.g., chloroplasts, mitochondria, the endoplasmic reticulum, the Golgi apparatus, and lysosomes) are the structural components of eukaryotic cells and carry out the metabolic processes of eukaryotic cells (Table S7). Similarly, organs serve their organism and multiple organs collaborate to carry out the metabolic activities of the organism (Kleine and Rossmanith, 2016). The living systems in group 1 are not self-sustaining entities (Table S5). For example, an arm depends on a person or a leaf on a plant. We therefore define the living systems in this cluster as the 'Component living system' (CS). The structure, process and function of each CS are controlled by its higher-level (parent) living system.

The shared features of the living systems in group 2 are high complexity in structure and process (Table S8). In general, the complexity of self-organized systems has several aspects. (1) The systems are composed of numerous components. These components together constitute the structure and processes of their parent system (Campeanu and Campeanu, 2012; Kulbacka et al., 2017). (2) The components in the system are closely interrelated. (3) The process is non-linear and it is difficult to infer the behavior of the whole from analysis its parts. (4) The system has the history that is irreversible, that is evolution. (5) New features of the systems continue to emerge in evolution. (6) The system has uncertainty and adaptability to some extent. All the living systems in this group meet the above six characteristics. For example, a eukaryotic cell consists of hundreds to thousands of organelles (Chang and Ge, 2005; Xu et al., 2020; Alberts et al., 2022). All the living systems in this group have a life history from birth to death. We define group 2 as 'Autonomous living system' (AS). It is worth noting that prokaryotic cells are also self-sustaining but lacks various components (organelles), and a distinct life history because they reproduce by binary fission. We therefore define prokaryotic cell as the 'pro-autonomous living system'.

The living systems in group 3 differ from groups 1 and 2 in structure and process. In group 3, a living system is an aggregate of the AS at the level immediately below (Table S5). For example, multiple proteins form a protein complex (Robinson et al., 2007), an adipose tissue contains many cells with similar morphology and function, and a human population is composed of many human individuals. We define living system in group 3 as the ‘Repeat-Autonomous living system’ (RAS). A population of multicellular organisms is an aggregate of conspecific organisms and one of a gene pool of a species, and it is the arena of intraspecific competition for resources, so it is an existing unit of a species of living system. Besides, populations exhibit typical structures and processes of RAS group (Table S9). Some models, such as the Logistic Model, competition model and cooperation model, share many traits between different levels of living systems in RAS. The significance of this group lies in: (1) resource acquisition and utilization, (2) reproductive units, (3) units for evolutionary selection.

Besides the shared features, new features emerge as level increases, with the size getting larger and information getting complex (Odum, 1959; O’Neill et al., 1989; Farnsworth et al., 2013; Cottam and Vounckx, 2021). The major differences between levels of living systems are at least five points: (1) Higher-level systems within a group have a larger size and contain more components that are multitudinous and processes than a lower level of living system (O’Neill et al., 1989). A system that is large size enough is the foundation for achieving complex systems and self-organization (Anderson, 1972). (2) Higher-level systems have a longer lifespan that raises the resources acquisition and utilization ability than lower level of living system (Pavé, 2006; West, 2017). A system with a sufficiently long time is also the foundation for achieving complex systems. (3) Higher-level systems have more complex regulation than lower-level systems. For instance, the unicellular eukaryote can only acquire intracellular regulations whereas hormonal and nervous control emerged in multicellular organisms in controlling and feedback (Chang et al., 2021). (4) Higher-level systems have a more stable internal state and resisting external disturbances than a lower-level system (Chang et al., 2021). For example, the thermal and chemical homeostasis as well as hydrostatic equilibrium and salt homeostasis exist in mammals and birds while only hydrostatic equilibrium and salt homeostasis in unicellular eukaryotes (Hohmann, 2002; Terrien et al., 2011; Wu et al., 2025). (5) Nevertheless, the organization of a higher level of living systems may be looser than that of a lower level of living system because it has had less time to develop. For instance, the structure and process of eukaryotes has become nearly complete over the past 1 billion years while that of eukarcities is still in developing over the past 300 years (Chang et al., 2021, 2022; Shang et al., 2024; Wu et al., 2025).

Until now, the 2D model for the hierarchical living systems has been built. The model can scale many qualitative features among the levels within a same group (Table 1-3, S5). However, the 2D model cannot link the living systems among the groups of the hierarchical living systems.

### 3.3 A three-dimensional model integrating the 1D and 2D models

Due to the cross directions of 1D and 2D models in scaling, we re-configure the groups in the 2D model and link all levels of living systems by the 1D model into a 3D helical model, namely HELIX (Fig. 3a). The belt is the 1D, which is transform of the lines of Fig. 1, and the three groups, CS, AS and RAS, appear in turn to compose a period. HELIX comprises 4 periods. Within each group, the quantitative traits in multiple periods (*Tqt*) scale with period as follows,

$$Tqt_{p,g} = Tqt_{0,g} \cdot {\lambda_g}^p \tag{3}$$

where $Tqt_{p,g}$ is the quantitative trait (length, volume, mass and time) of living systems at period $p$ in group $g$ (p = 1–4; 1–3 for RAS), $Tqt_{0,g}$ is the extrapolated value at period zero, estimated as the intercept of the log-linear regression, $\lambda_g$ is the rate of increase per period in a group $g$, representing the multiplicative factor by which the quantitative trait increases per period.

In the periodic scaling models in 3D, all the regressions of the quantitative traits are significant except those for mass and active duration in RAS (Fig. 3b-e). The $\lambda_g$ of length in response to period number of the three groups varies from $10^{3.7}$ to $10^{5.2}$, and the exponents are approximately 3- to 4-fold of that at the level of living systems in 1D. The exponents of the λ of volume, mass and time are also approximately 3- to 4-fold of their corresponding 1D value. Similar to the second-class structure model of protein, HELIX describes the ‘second-class structure’ of the hierarchy of living systems on the Earth.

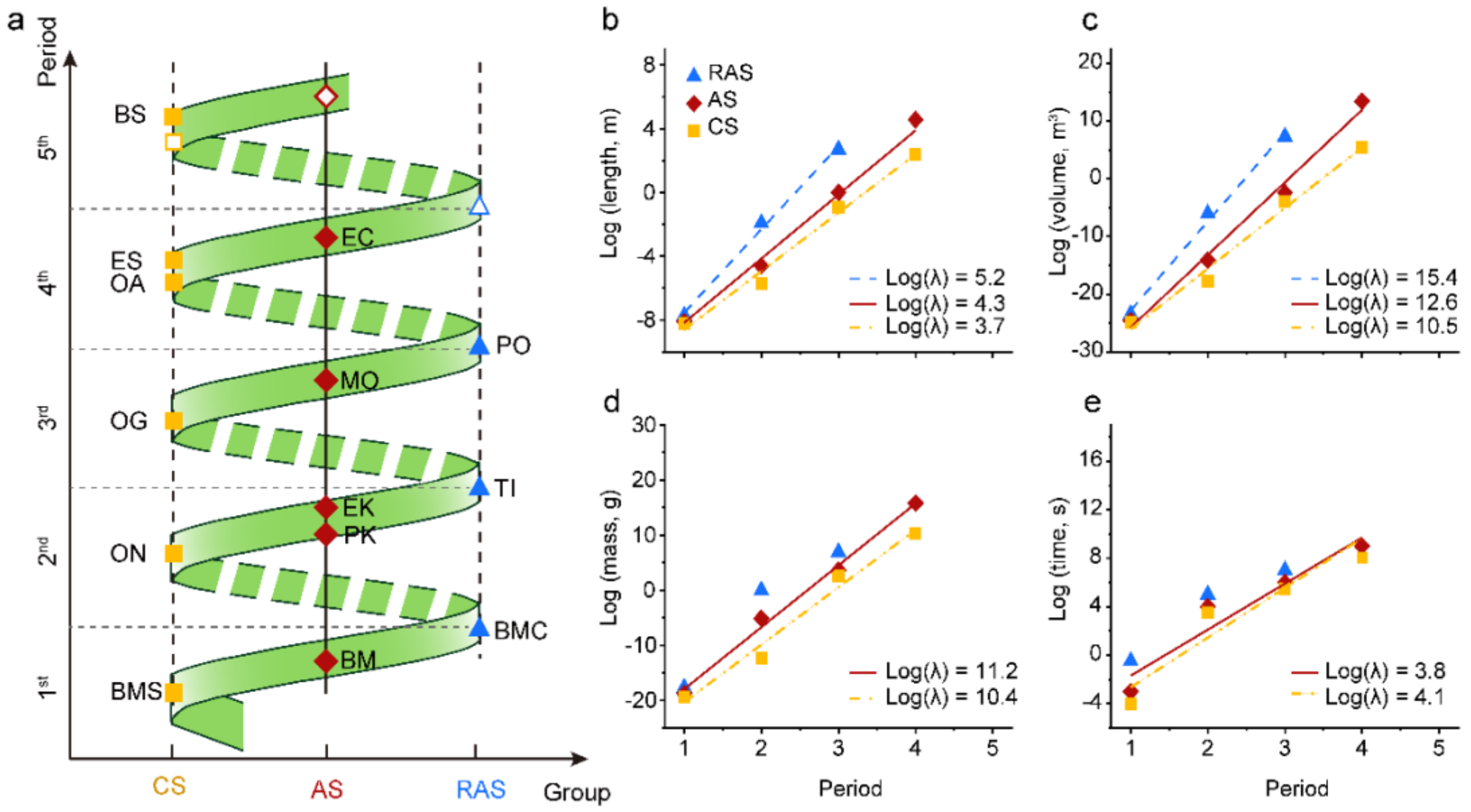

**Fig. 3 HELIX model of the hierarchical living systems and scaling of the groups.** (a) The HELIX model with three groups, CS, AS, and RAS linked by the periods, the living systems in known 4 periods is presented by closed points, and those in a potential 5$^{th}$ period is presented by open points. (b-e) Scale of the quantitative traits of living systems of each group, (b) length ($R^2$ of CS, AS, RAS is 0.99, 1.00, 1.00 respectively), (c) volume ($R^2$ is 0.99, 0.99, 1.00, respectively), (d) mass (fresh weight, $R^2$ of CS, AS is 0.98, 0.99 respectively), (e) time ($R^2$ of CS, AS is 0.93, 0.93, respectively). Abbreviations see Fig. 1.

HELIX scales both the quantitative and qualitative features of the 14 levels of living systems at adjacent and non-adjacent levels. It solves a difficulty of existing hierarchical models, none of which can scale across all levels. In this sense, HELIX describes the 'second-class structure' of the hierarchy of living systems. Beyond scaling, HELIX is compatible with many present theories, principles and models that for a few cross-level living systems (Supplementary Text S6). For example, in the AS group, homeostasis, originally defined for the physiology of organisms (Cannon, 1929), is shared by all autonomous systems as a product of self-organization. In cities it appears as stable supplies of water, food, power and light, and as reduced temperature variation (Chang et al., 2021). The three fusions of endosymbiosis (Margulis, 2010) can all be explained as origin mechanisms of autonomous living systems in AS groups. At the global scale, GAIA (Lovelock and Margulis, 1974) is the self-organizing mechanism that maintains the homeostasis of the Earth in AS. Another example is that CS provide services to their parent systems in AS group and link CS with AS groups. For example, ecosystem services (Costanza et al., 1997) in city level are provided by artificial and natural ecosystems (Chang et al., 2021). The living systems in AS aggregate into RAS by information, including gene, neural signals and hormone, and the human information such as meme (Dawkins, 1976). Analogous to genomics in biological systems, the new field 'Culturomics' enables the systematic study of cultural information, including its frequency, mutation rate and evolutionary dynamics over time (Perreault, 2012; Newberry et al., 2017; Hou et al., 2025). See Supplementary Text S6 for details.

**4 The newly emerged levels of living systems predicted by HELIX**

In addition to the explanation, HELIX suggests that some unknown living systems may exist. In Fig. 3a, HELIX leaves three positions unoccupied: (1) there should be an unknown living system larger than biosphere, and located in the AS group in a higher period than eukarcity (hereafter unknown 1); (2) there should be an unknown human-dominated living system in the CS group in a higher period other than the biosphere (hereafter unknown 2); (3) there should be an unknown living system larger than eukarcity in the RAS group in a higher period (hereafter unknown 3). Then, we use the level scaling models in 1D (Eq. 1) and the periodic scaling models in 3D (Eq. 3) to predict the size of these unknown systems (Table 1).

**Table 1. Validation of the 1D and 3D models against observed length, mass, and volume of the newly emerged unknown living systems.**

| System | Prediction by 1D, level scaling [95% PI] | Prediction by 3D, periodic scaling [95% PI] | Observed |
|---|---|---|---|
| $Log_{10}$ length (m) | | | |
| Unknown 1 | 9.34 [7.06, 11.62] | 8.61 [5.74, 11.48] | **7.82** |
| Unknown 2 | 7.01 [4.87, 9.15] | 6.05 [1.40, 10.70] | **6.20** |
| Unknown 3 | 5.85 [3.77, 7.93] | 8.11 [-3.04, 19.26] | **5.50** |
| $Log_{10}$ mass (g) | | | |
| Unknown 1 | 29.42 [24.52, 34.33] | 26.96 [17.24, 36.69] | **24.15** |
| Unknown 2 | 23.04[18.45, 27.64] | 21.40 [5.17, 37.63] | **21.76** |
| Unknown 3 | 19.85 [15.39, 24.31] | – | **17.26** |
| $Log_{10}$ volume ($m^3$) | | | |
| Unknown 1 | 26.78 [19.81, 33.75] | 24.52 [11.11, 37.93] | **22.24** |
| Unknown 2 | 19.92 [13.39, 26.45] | 16.43 [1.17, 31.70] | **16.62** |
| Unknown 3 | 16.49 [10.15, 22.83] | 23.36 [-15.05, 61.76] | **14.93** |

### 4.1 The unknown living system 1

Unknown 1 should be the highest level of living system in the hierarchy. The prediction of the length by 1D model is ~$10^{9.3}$ m while by 3D is ~$10^{8.6}$ m (Table 1). Humans have expanded the size of the Earth by artificial satellites (Chang and Ge, 2017). More than 20,000 artificial satellites currently operate in outer space above ground surface (Fig. 4a, b). This places the upper boundary at about $3.58 \times 10^7$ m above the Earth's surface (Teunissen and Montenbruck, 2017). The extension of the lower boundary of the Earth by human activities remains on the order of $10^4$ m below the surface. Now the diameter (length) of the human modified Earth has been expanded to $10^{7.82}$ m (Table 1). Human activities make the Earth system much larger than biosphere, coupling the biosphere with other spheres through artificial mass, energy and human information. We define this system as the 'Anthropogenic Earth Living System', or anthro-Earth (AE).

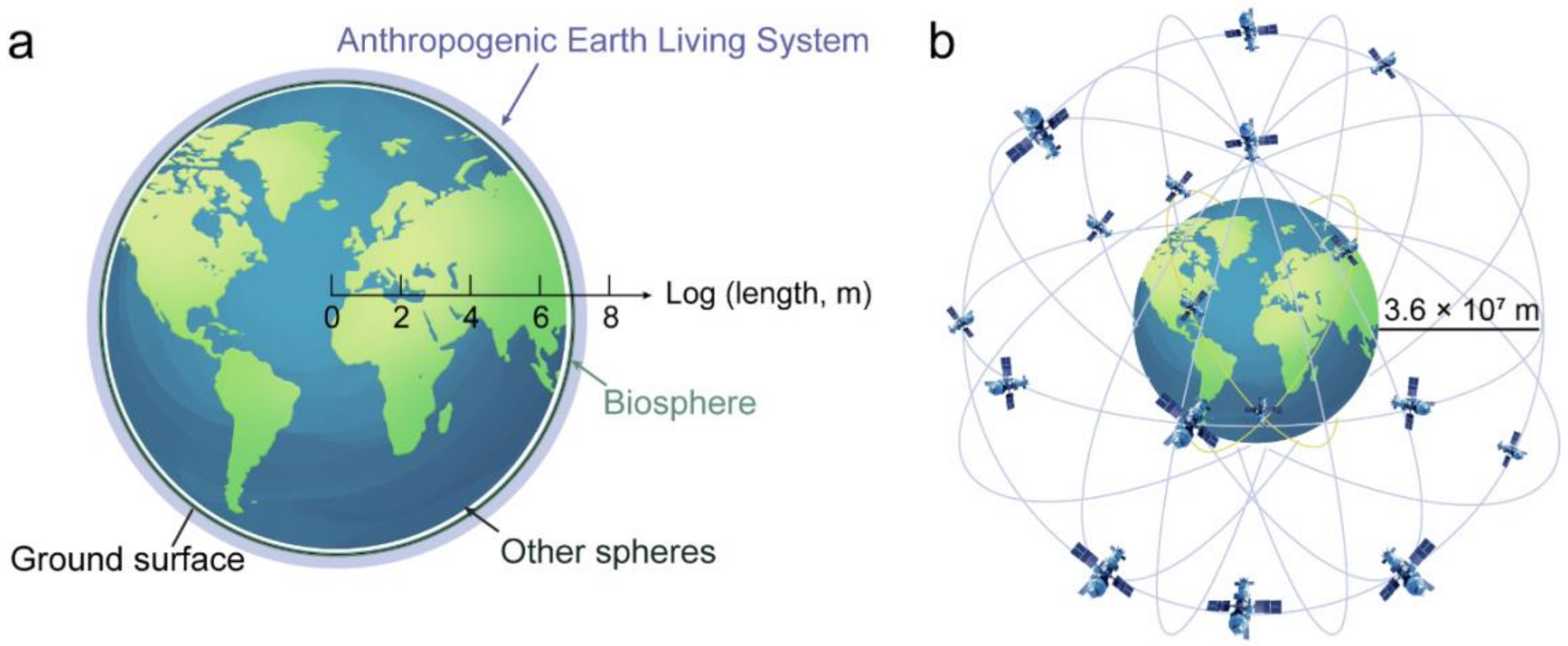

**Fig. 4 Illustration for the diameter of the Anthropogenic Earth Living system.** (a) Biosphere is only one thin sphere on the Earth surface and it co-exist with atmosphere, geosphere, pedosphere and hydrosphere, as well as artificial satellite sphere; (b) the extension scale of artificial satellite above Earth surface to be the out boundary of the Anthropogenic Earth living system.

To verify the quantitative prediction, we calculated the length, mass and volume of the Anthro-Earth (Supplementary Text S7). The length of Anthro-Earth is the diameter, defined as twice the distance from the Earth's center to the outer boundary of the artificial satellite sphere. The mass of Anthro-Earth is the sum of the masses of the biosphere, hydrosphere, pedosphere, and anthropogenic mass. The observed length ($10^{7.82}$ m) and volume ($10^{22.2}$ $m^3$) of AE fall within the 95% prediction intervals of both models, whereas its mass ($10^{24.15}$g) falls below the 1D interval but within the 3D interval (Table 1).

Beside the size, human-made materials increased continuously since industrialization. The human-made material in the past ~300 years exceeded the living biomass (wet) around 2020 ± 6 year (Elhacham et al., 2020). Meanwhile, human culture, including science, technology, institutions and art, is gradually becoming the major regulator of the Earth (Chang and Ge, 2005; Kuwae et al., 2024). Human culture forces the evolution on the Earth increasingly and pushes the Earth into the Anthropocene (Steffen et al., 2015; Lehman et al., 2021). Humanity has reinforced this regulation into an Anthropogenic GAIA (AGAIA), a new control–feedback mechanism for the anthro-biogeochemical cycles (Chang and Ge, 2005). AE is a life–culture system based on the biosphere and coupling all spheres of the Earth, with human as its regulatory center (Chang and Ge, 2005, 2010). Its subsystems are coordinated through near-real-time telecommunication, much as the nervous system coordinates an organism. In HELIX, the highest newly emerged level of living system should be located in the AS group (Fig. 2, 3). AE shows the traits of this group in three aspects: artificial infrastructure, energy and information in Supplementary Text S8.

The concept of AE overlaps to some extent with many present concepts such as technosphere, noosphere, Anthropocene, GAIA, planetary social-ecological system. For example, technosphere only describes the global assemblage of human-created buildings, infrastructure, machinery and artifacts which plays a major role in present-day Earth-system dynamics; the Anthropocene mainly emphasizes the human dominated evolutionary stage. AE integrates all these aspects and ideas, and it is a physical entity occupying three dimensions of space and one dimension of time since industrialization. Its structure consists of the planet, the biosphere, and several other spheres, as well as the artificial matter sphere (including satellites). It also emphasizes the relationships among various components, including the technological sphere and cognitive sphere coupled with GAIA mechanism.

### 4.2 The unknown living system 2

By HELIX, we predict that unknown 2 should be the component of the anthro-Earth so it should occupy the CS position (Figs. 2 and 3). The size of this living system is lower than the anthro-Earth (Table 1). We propose it should be country or nation. The length of a country is the diameter of a circle with the same area as the country. Its volume is that of a cylinder with this base and a height spanning the vertical extent of the biosphere (~$10^4$ m above and below the ground surface). Its mass (wet weight) is the mass of AE apportioned by the country's share of the global land area (Supplementary Text S6). The resulting length, volume and mass of a country are $10^{6.2}$ m, $10^{16.6}$ $m^3$ and $10^{21.78}$ g, respectively, all within the 95% prediction intervals of both the 1D and 3D models (Table 1).

A country (or nation) is a political and cultural entity situated on a specific territory (World Bank, 2025), and the function components of AE (Fig. S4). For example, the large volume and global circulation of goods and services have led to close interdependence among countries worldwide (Cumming et al., 2014; Chang et al., 2021). The anthro-biogeochemical cycles are organized mainly by countries (international trade and trade barriers). For example, international trade tightly links all countries (Kiljunen, 1989), and no country is self-sufficient in the world. The trade frequency between large countries, such as China and USA, is as high as $10^4$ to $10^5$ trades each year (Ma et al., 2006; WTO, 2023). The qualitative traits fit the prediction of the 3D model. This interaction extends to the dissemination, collaboration, competition, and mutual coupling in the realms of material, energy, and so on.

In addition, frequent exchanges of anthropogenic materials, energy and information, together with global institutions and agreements such as the 'United Nation', 'Food and Agriculture Organization (FAO)', 'Kyoto Protocol for Global Climate Change', 'Convention on Biological Diversity', link countries into a unit. We define a country as an 'Organmega', meaning a 'great scale organ' of anthro-Earth. In the 1D model of hierarchical structure, an organmega is not only a subunit of anthro-Earth but is itself composed of an unknown system made up of eukarcities, together with a few regions not influenced by humanity, such as protection area (Chang et al., 2021). Organmega share the qualitative traits of CS (Table S7). In particular, an organmega sustains itself through politics, institutions, and cultures, while being subject to economic, ecology, and cultural influences. Therefore, an organmega is the semi-autonomous in its development, birth and death. Meanwhile, other countries also influence these processes. The intergenerational transmission of the culture of a country is by means of the semi-autonomous manner, which is similar to organelles of eukaryotic cells. The periodic position of the organmega fits the 3D helical curve and the regression model of the HELIX.

**4.3 The unknown living system 3**

HELIX predicts that unknown 3 occupies the RAS position (Fig. S8), with a size between those of the eukarcity and the organmega. Each RAS level is an aggregate of the AS level immediately below it, so Unknown 3 should be an aggregate of eukarcities. We propose the city cluster (Dash Nelson and Rae, 2016; Fang and Yu, 2017; Shen et al., 2022) as a candidate. Following the boundary rules for eukarcities, the length of a city cluster is the diameter of a circle with its mean area; its volume is that of a cylinder extending from 10,000 m below to 1,000 m above the ground surface; and its mass is the mean eukarcity mass scaled by the ratio of city-cluster area to eukarcity area (Supplementary Text S6).

We compiled 13 city clusters from China, the United States and seven other countries, each delimited by government planning documents or official statistics. According to the dataset that we collect worldwide, the average length, volume and mass are $10^{5.5}$ m, $10^{14.9}$ $m^3$ and $10^{17.3}$g, respectively. All three fall within the 95% prediction intervals of the 1D model. In the 3D model, length falls within the interval, whereas volume falls just below its lower bound; mass could not be tested because the within-group regression for mass in the RAS group is not significant (Table 1). Future studies should pay more caution on the development of this level of living system.

### 4.4 Verify the three newly emerged living systems by 2D

To test whether the three candidates also share the qualitative traits of their predicted groups, we scored AE, the country (CT) and the city cluster (CC) on the 16 traits (Table S4) and repeated the cluster analysis (2D) for all 17 levels. The memberships of the 14 known levels were unchanged, and each candidate joined the group predicted by HELIX: AE clustered with the AS, CT with the CS, and CC with the RAS (Fig. S5). As recently emerged systems, however, the country and the city cluster are still developing in boundaries and functions and need further study.

## 5 The updated HELIX including the newly emerged living systems

The three new emerged living systems (or quasi-living systems) add a period to the former HELIX (Fig. 3) and then the updated HELIX model contains five periods (Fig. 5a). The quantitative traits of the length, volume and mass in respond to level and period keep the λ not significant change (Fig. S2, Fig. 5b). Compare with former HELIX, the λ of the length, volume, and mass of the living systems in response to level decreased from $10^{1.16}$ to $10^{1.10}$, $10^{3.43}$ to $10^{3.20}$, and $10^{3.19}$ to $10^{2.90}$, respectively (Fig. S2), while $R^2$ remained high (0.97–0.98), and in response to periods keep similar and a little bit lower. The 2D model confirms that the three groups remain stable when add the predicted living systems (Fig. S5). The 3D model confirms that the HELIX remains robust after adding the fifth period of the HELIX. It validates the legitimacy of the existence of these "quasi-living systems" or living systems.

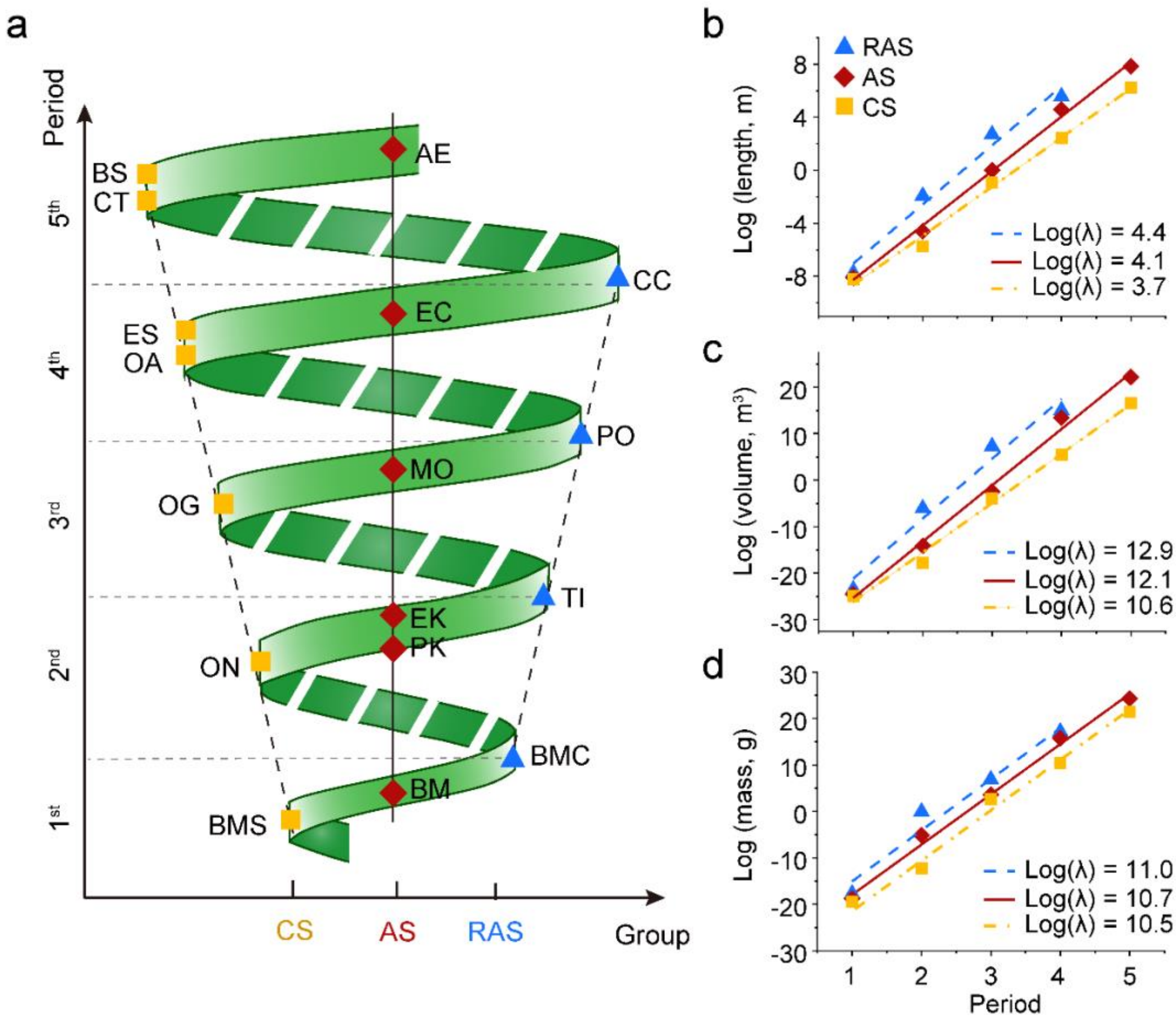

**Fig. 5 Updated HELIX model including all the hierarchical living system up to now.** (a) Helical model, the belts include most living systems except a few, red point-autonomous living systems, yellow point-component living systems, blue point-repeat autonomous living system, prokaryotic cells and natural ecosystems are quasi-perfect systems so they are not in the main

order, and the extending in scale from below period to above period illustrates the scaling of the quantitative traits. (b-d) Scale of each of the three groups in response of period, (b) length ($R^2$ of CS, AS, RAS is 0.99, 1.00, 0.98, respectively), (c) volume ($R^2$ is 0.99, 0.99, 0.97, respectively), (d) mass ($R^2$ is 0.99, 0.99, 0.96, respectively). Abbreviations and order: the levels of 1-13 is the same with Figure 1, level 14 replaced to CC, city cluster, 15 CT, country, 16 BS, biosphere, 17 AE, Anthropogenic Earth living system. All regressions in (b-d) are $p < 0.01$. Compare with the 14-level HELIX, the mass of RAS to be significant.

## 6 The evolutionary mechanism of the hierarchy of living system uncovered by HELIX

Symbiosis theory focuses on the emergence of the living systems in AS group while the HELIX addresses the emergence of all three groups and the relationships among them. The 1st emergence of living systems includes period 1, BM, BM subunit and BM complex are the pre-living systems, and the emergence of prokaryote cell, which corresponds to the 1st fusion of symbiosis theory (Margulis, 2010). The 1st emergence began about 3.6 billion years ago, with the formation of pre-living materials, and lasted until the origin of life in the prokaryotic cell. Some biomacromolecules, such as RNA and DNA, are themselves information carriers, and the prokaryotic cell enclosed this genetic information within its outer membrane. The outer membrane separated the internal structure from the external environment, so salt homeostasis formed. Prokaryotes contributed greatly to GAIA, forming the aerobic atmosphere through more than 2 billion years of oxygen accumulation (Lovelock and Margulis, 1974). The material of this period was only carbon-based and consisted only of light elements of the Periodic Table of Elements. Energy was obtained from chemical sources by chemotrophy and, mainly, from the sun by photosynthesis.

The 2nd emergence of living systems in evolution covers period 2 and 3 of HELIX. After a long evolution history, the eukaryotic cell (in AS) emerged through symbiosis, corresponding to the 2nd fusion of symbiosis theory: some prokaryotic cells evolved into organelles (in CS), eukaryotic cells aggregated into tissue (in RAS). Then the multicellular organism (in AS) emerged soon and it correspond to the 3rd fusion of symbiosis theory (Margulis, 2010). Meanwhile, organs (in CS) and populations (in RAS) also emerged, forming the period 3 of HELIX. Multicellular organisms triggered a boom in biodiversity. Animals, plants and fungi diversified into kingdoms, phyla, classes, orders, families, genera, and species, with numerous variations below the species level. Food web and ecosystems also emerged. Genetic information remained the main heritable information, while multicellular organisms added hormones and nerves to transmit information and regulate activities among cells and organs (Chang and Ge, 2010). The material component of these two periods is carbon-based, with calcium added in animal skeletons. It is also limited to light chemical elements. Energy came mainly from photosynthesis.

The 3rd emergence in evolution spans periods 4 and 5 in the renewed HELIX. After *Homo* emerged ~2 million years ago, human language developed ~0.1 million years ago. Over the past 300 years, the Industrial Revolution (IR) gave rise to the eukarcity (in AS), while organara (in CS) and city cluster (in RAS) emerged at the same time, forming period 4 of HELIX. A new level of homeostasis formed within the eukarcity, especially in urban areas. From the 1950s, the 3rd IR gave rise to artificial satellites, which accelerated global integration through long-distance communication, and the anthro-Earth (in AS) emerged. AGAIA is the self-organization mechanism for homeostasis at the Earth level. Living matter in periods 4 and 5 is carbon-silicon based, that is, silicon dominates both city structure and information processing by computers, telecommunication and artificial intelligence. Energy now also comes from fossil fuels and nuclear power. A new type of storable and heritable information, human cultural information, emerged in this period. Chang and Ge (2005) divided it into cene and tene. Cene is the formal cultural information encoded in binary code (0 and 1), such as scientific, technical and policy texts. Its binary code corresponds to the nucleotide bases of DNA, and it is migrating from carbon-based media, such as books and maps, to silicon-based digital storage (Bentley and O'Brien, 2012). Currently, information units are named "Token" in data computation and trading. In contrast, tene consists of small snippets frequently transcribed from cene, such as oral messages and small files, which guide most daily activities. It is similar to the RNA in cells.

Although higher levels have emerged, evolution is conservative in mechanism. Cultural information flows from cene to tene and then to social activities and artificial products, paralleling the central dogma from DNA to RNA to protein (Crick, 1970). Individuals internalize cultural information through teaching and imitation, and turn it into knowledge and skills that guide behaviour (Mesoudi, 2017; Creanza et al., 2017). The reverse flow also occurs: innovations recorded in language are written back into cene as new heritable information (Lansdall-Welfare et al., 2017; Newberry et al., 2017). Like genes, cenes replicate and spread with high fidelity, and can now be analysed at scale (Lansdall-Welfare et al., 2017; Baumard et al., 2022; Novoa et al., 2025). Analogous to genomics, culturomics, which has emerged since 2011, enables the systematic study of cenes, including their frequencies, mutation rates and evolutionary dynamics (Michel et al., 2011; Perreault, 2012; Newberry et al., 2017; Hou et al., 2025). At the levels of the eukarcity and the anthro-Earth, electronic communication acts like nerves and flows of energy, goods and money act like hormones. Artificial intelligence accelerates the transcription of cene into tene and its translation into action, reinforcing cene as a regulator of the anthro-Earth.

In summary, when a new living system in AS emerges, in most cases through fusion, its components become the CS of that period, as organelles did in the eukaryotic cell and organaras in the eukarcity. Its aggregates form the living system in RAS, such as tissues of eukaryotic cells and populations of multicellular organisms. In this way, a complete period is added to HELIX. Each emergence rests on an expansion in matter, energy and information. Of the three, information is the leading dimension, because it stores, transmits and regulates the structures built from matter and energy. The evolution of life is thus the evolution of information. This central role of information places the AS group at the center of HELIX. Life originated with the prokaryotic cell in AS, and every major upgrade of life, from the eukaryotic cell and the multicellular organism to the eukarcity and the anthro-Earth, occurred in AS.

## 7 Conclusion

For a long time, the development of ecology, geology, genetics, and evolutionary science has involved multiple levels of living systems. In recent decades, advances in systems science, information science and culturomics have provided theories and tools that cover several levels. However, a universal model that spans all levels is still lacking. Based on interdisciplinary assumptions, knowledge, and tools, we established HELIX, which integrates two models. Fitted to the 13 known levels, the one-dimensional model describes exponential relationships between quantitative traits (length, volume, mass and lifespan) and level, and power-law relationships among these traits, allowing the traits to be estimated across levels. The two-dimensional model clusters the 14 levels into three groups by 16 traits of structure, process, and function, and the living systems within each group share many characteristics. HELIX combines the one- and two-dimensional models in three dimensions, arranging the levels along a helical belt. It thus retains the scaling across levels and captures the periodic recurrence of the qualitative traits of the three groups. HELIX is compatible with many existing theories, such as GAIA, endosymbiosis theory, homeostasis, and meme hypothesis, and may serve as a theoretical framework for a new synthesis of ecology, geology, genetics, developmental biology and evolutionary science. From the perspectives of matter, energy and information, HELIX provides an organization mechanism for the hierarchical spectrum of living systems. Moreover, HELIX has predictive capabilities: it predicts three newly emerged candidate levels, estimates their scale and assigns them to groups. In particular, with human culture integrated into living systems and human information now dominant, we hypothesize that the Anthropogenic Earth has developed into the highest level of living systems in the hierarchy. In the future, other levels may be recognized. HELIX may thus promote cross-level and interdisciplinary studies of living systems.

**Declaration of generative AI and AI-assisted technologies in the writing process**

During the preparation of this work, the authors used Claude (Anthropic) in order to improve the language and readability of the manuscript. After using this tool, the authors reviewed and edited the content as needed and take full responsibility for the content of the publication.

**Supplementary materials for**

# A predictive scaling framework for hierarchical living systems: From biomacromolecule to the Anthropogenic Earth system

**This PDF file includes:**

Table S1 An evaluation of organaras in a eukarcity

Table S2 Traits of the 14 hierarchical living systems and Ontology ID

Table S3 Ontologies used to define and score traits at each of the 14 known levels of living systems

Table S4 Scoring criteria for traits of living systems used in cluster analysis

Table S5 Qualitative trait scores of the 14 hierarchical living systems used for cluster analysis

Table S6 Quantitative traits of scale of the 14 known levels of living systems

Table S7 Qualitative trait of the living systems in Group Component living system

Table S8 Qualitative trait of the living systems in Group Autonomous living system

Table S9 Qualitative trait of the living systems in Group Repeat-Autonomous living system

**Supplementary text**

**Text S1 Data collection and calculation for size of the 14 levels of living systems**

**(1) Biomacromolecule subunit**

A biomacromolecule subunit refers to a single, independent polypeptide chain within a macromolecular structure such as a protein or nucleic acid (Steven et al., 1983).

**Volume** The volume of a biomacromolecule subunit ($V_{bms}$) is,

$$V_{bms} = \frac{M_{bms} \times V_{bm}}{M_{bm}} \tag{S1}$$

where $M_{bms}$ is the average molar mass of the subunit; $M_{bm}$ is the molar mass of the biomacromolecule, taken from Clogston and Caffrey (2005); $V_{bm}$ is the volume of the biomacromolecule.

**Length** In this study, the biomacromolecule subunit is treated as a sphere, and its length ($d_{bms}$) is,

$$d_{bms} = 2 \times \sqrt[3]{\frac{3 \times V_{bms}}{4 \times \pi}} \tag{S2}$$

where π is taken as 3.14.

**Mass** Since molar mass is the total mass of $6.02 \times 10^{23}$ molecules, the mass of the subunit ($m_{bms}$) is,

$$m_{bms} = \frac{M_{bms}}{N_A} \tag{S3}$$

where $M_{bms}$ is the molar mass of the subunit, taken from Steven et al. (1983); $N_A$ is Avogadro's constant, taken as $6.02 \times 10^{23}$.

**(2) Biomacromolecule**

A biomacromolecule refers to a large polymeric compound formed by the covalent linkage of many atoms or smaller molecules (Alberts et al., 2022). The major biomacromolecules include proteins, nucleic acids (DNA and RNA), polysaccharides (such as starch and cellulose), and certain complex lipid components (such as phospholipids).

**Volume** The volume of a biomacromolecule ($V_{bm}$) is taken from Clogston and Caffrey (2005),

$$V_{bm} = \frac{\sum_{i=1}^{N} V_{bmi}}{N} \tag{S4}$$

where $V_{bmi}$ is the volume of the *i-th* biomacromolecule, $N$ is sample size.

**Length** The biomacromolecule is treated as a sphere, and its length ($d_{bm}$) is

$$d_{bm} = 2 \times \sqrt[3]{\frac{3 \times V_{bm}}{4 \times \pi}} \tag{S5}$$

where $\pi$ is taken as 3.14.

**Mass** The mass of the biomacromolecule ($m_{bm}$) is

$$m_{bm} = \frac{M_{bm}}{N_A} \tag{S6}$$

where $M_{bm}$ is the molar mass of the biomacromolecule, taken from Clogston and Caffrey (2005); $N_A$ is Avogadro's constant, taken as $6.02 \times 10^{23}$.

**(3) Biomacromolecular complex**

A biomacromolecular complex is an assembly of multiple biomacromolecules held together by non-covalent bonds or physical interactions, forming a structurally and functionally integrated complex that cooperatively carries out a series of sequential biochemical reactions, such as the RNA exosome, the proteasome, and the nuclear pore complex (Robinson et al., 2007).

**Volume** The volume of a biomacromolecular complex ($V_{bmc}$) is converted from mass and calculated as:

$$V_{bmc} = \frac{M_{bmc} \times V_{bm}}{M_{bm}} \tag{S7}$$

where $M_{bmc}$ is the average molar mass of the biomacromolecular complex; $M_{bm}$ is the molar mass of the biomacromolecule, taken from Clogston and Caffrey (2005); and $V_{bm}$ is the volume of the biomacromolecule.

**Length** The biomacromolecular complex is treated as a sphere, and its length ($d_{bmc}$) is

$$d_{bmc} = 2 \times \sqrt[3]{\frac{3 \times V_{bmc}}{4 \times \pi}} \tag{S8}$$

where $\pi$ is taken as 3.14.

**Mass** The mass of the biomacromolecular complex, $m_{bmc}$ is calculated as:

$$m_{bmc} = \frac{M_{bmc}}{N_A} \tag{S9}$$

where $M_{bmc}$ is the molar mass of the biomacromolecular complex, taken from Brindley et al. (1969); $N_A$ is Avogadro's constant, representing the number of molecules per mole, taken as $6.02 \times 10^{23}$.

**(4) Prokaryote cell**

The prokaryotic cell is the most primitive type of cell, lacking a well-defined nucleus and a nucleolus (Alberts et al., 2022).

**Length** The radius of the prokaryotic cell, $r_{pk}$, is taken as 2.5 μm, following the estimate of Alberts et al. (2022); the length $d_{pk}$ is therefore 5 μm ($5\times10^{-6}$ m).

**Volume** The prokaryotic cell is treated as a sphere, and its volume ($V_{pk}$) is

$$V_{pk} = \frac{4 \times \pi \times {r_{pk}}^3}{3} \tag{S10}$$

where π is taken as 3.14.

**Mass** The mass of the prokaryotic cell($m_{pk}$) is taken form DeLong et al. (2010),

$$m_{pk} = \frac{\sum_{i=1}^{N} m_{pki}}{N} \tag{S11}$$

where $m_{pki}$ is the mass of the i-th prokaryotic cell, $N$ is sample size.

**(5) Organelle of eukaryotic cell**

Organelles are structures within eukaryotic cells that have distinct morphological features and perform specific functions, such as plastids, mitochondria, ribosomes, chloroplasts, and spherosomes (Alberts et al., 2022).

**Volume** The volume of an organelle ($V_{on}$) is taken from Grimes et al. (1974),

$$V_{on} = \frac{\sum_{i=1}^{N} V_{oni}}{N} \tag{S12}$$

where $V_{oni}$ is the volume of the *i-th* organelle, $N$ is sample size.

**Length** The organelle is treated as a sphere, and its length ($d_{on}$) is

$$d_{on} = 2 \times \sqrt[3]{\frac{3 \times V_{on}}{4 \times \pi}} \tag{S13}$$

where π is taken as 3.14.

**Mass** The mass of an organelle ($m_{on}$) is estimated from the literature (Glas and Bahr, 1966; Bradley and Quarmby, 2005; Takamori et al., 2006), including the mitochondria, ribosomes, and endoplasmic reticulum of eukaryotic cells, as well as the flagella of prokaryotic cells,

$$m_{on} = \frac{\sum_{i=1}^{N} m_{oni}}{N} \quad \text{(S14)}$$

where $m_{oni}$ is the mass of the i-th organelle, $N$ is sample size.

**(6) Eukaryote cell**

Eukaryote cells are cells with a nucleus enclosed by a nuclear membrane, possessing complex structures and diverse functions, and serve as the fundamental units of all eukaryotic organisms (Alberts et al., 2022).

**Length** The length of the eukaryotic cell is taken as 25 μm, following the estimate of Alberts et al. (2022).

**Volume** The volume of a eukaryotic cell ( $V_{ek}$) is taken from literature (Grimes et al., 1974; Bradley and Quarmby, 2005; Hoang et al., 2019),

$$V_{ek} = \frac{\sum_{i=1}^{N} V_{eki}}{N} \quad \text{(S15)}$$

where $V_{ek}$ is the volume of the i-th eukaryotic cell, and $N$ is sample size.

**Mass** The mass of a eukaryotic cell ($m_{ek}$) is taken form DeLong et al. (2010),

$$m_{ek} = \frac{\sum_{i=1}^{N} m_{eki}}{N} \quad \text{(S16)}$$

where $m_{ek}$ is the mass of the i-th eukaryotic cell, and $N$ is sample size.

**(7) Tissue of organism**

Tissue refers to a group of cells with similar morphology and related functions, together with the extracellular matrix, that are integrated through cell–cell connections and matrix interactions to perform specific functions (Cheong et al., 1990). Tissue represents the next level of organs.

**Length** In this study, the tissue is treated as a sphere, and its length ($d_{ti}$) is

$$d_{ti} = 2 \times \sqrt[3]{\frac{3 \times V_{ti}}{4 \times \pi}} \quad \text{(S17)}$$

**Volume** The volume of the tissue ($V_{ti}$) is

$$V_{ti} = \frac{m_{ti}}{\rho_{ti}} \quad \text{(S18)}$$

where $\rho_{ti}$ is the density of the organism, taken as 1.02 g/cm$^3$.

**Mass** The mass of tissues ($m_{ti}$) is estimated following Cheong et al. (1990), based on the mean mass of epithelial, connective, muscle, and nervous tissues reported for human, mouse, and other samples,

$$m_{ti} = \frac{\sum_{i=1}^{N} m_{tii}}{N} \tag{S19}$$

where $m_{ti}$ is the mass of the *i-th* tissue, and $N$ is sample size.

**(8) Organ of organism**

An organ is a substructure of a multicellular organism with a distinct morphology that performs specific functions to support the physiological processes of the organism (Kim et al., 1983). In animals, organs include the heart, eyes, and ears, with approximately 78 types in total. In plants, there are six major types of organs: roots, stems, leaves, flowers, fruits, and seeds.

**Volume** The organ volume ($V_{og}$) is averaged by combining data from literature (Luccichenti et al., 2003; Geraghty et al., 2004; Jia et al., 2005),

$$V_{og} = \frac{\sum_{i=1}^{N} V_{ogi}}{N} \tag{S20}$$

where $V_{og}$ is the volume of the *i-th* organ, and $N$ is sample size.

**Length** In this study, the organ is treated as a sphere, and its length ($d_{og}$) is

$$d_{og} = 2 \times \sqrt[3]{\frac{3 \times V_{og}}{4 \times \pi}} \tag{S21}$$

where π is taken as 3.14.

**Mass** The mass of organs ($m_{og}$) in this study is obtained from the literature and calculated as the mean mass of 15 organs, including the heart, liver, spleen, lungs, and kidneys (Kim et al., 1983; ICRP, 2003; Tanaka et al., 1998),

$$m_{og} = \frac{\sum_{i=1}^{N} m_{ogi}}{N} \tag{S22}$$

where $m_{og}$ is the mass of the *i-th* organ, and $N$ is sample size.

**(9) Multicellular organism**

Multicellular organism refers to a living system that have relatively independent structures, respond to external stimuli, and function independently (Pavé, 2006; Chang and Ge, 2010).

**Length** The length of the multicellular organism is taken as 1 m, following the estimate of Pavé (2006).

**Mass** The mass of organisms in this study was taken as the average of the mass of mammals, birds, reptiles, amphibians, fish, and invertebrates (Hatton et al., 2019; West et al., 2002),

$$m_{mo} = \frac{\sum_{i=1}^{N} m_{moi}}{N} \tag{S23}$$

where $m_{mo}$ is the mass of the i-th multicellular organism, and $N$ is sample size.

**Volume** The volume of the multicellular organism ($V_{mo}$) is calculated as:

$$V_{mo} = \frac{m_{mo}}{\rho_{mo}} \tag{S24}$$

where $\rho_{mo}$ is the density of the organism, taken as 1.02 g/cm$^3$.

**(10) Population**

Population refers to a structural and functional unit consisting of a group of individuals of the same species in a certain time and space (Chang and Ge, 2010).

**Area** The area of a population ($S_{po}$) refers to the land area occupied by a specific species population within a given time and space (Anderson, 2022). In this study, the area of animal populations was derived from territory size during the breeding season, covering 19 species in total, including vertebrates such as mammals, birds, amphibians and reptiles, as well as invertebrates like ants and bees (Mosser and Packer, 2009; Chakanyuka and Utete, 2022; Zammarelli et al., 2024). We searched for animal population area on Web of Science, Google Scholar, PubMed and other databases using the search term “target species + territory”. For plant population area, land-cover data of angiosperms and gymnosperms derived from satellite imagery were retrieved. When searching for plant population area, we first searched for a specific plant species (e.g., pine), and then filtered results with satellite-image techniques such as Google Earth to obtain accurate outcomes. Finally, we screened the retrieved literature and obtained land-occupation data for seven plant populations.

$$S_{po} = \frac{\sum_{i=1}^{N} S_{poi}}{N} \tag{S25}$$

where $S_{poi}$ is the area of the i-th population, and $N$ is sample size.

**Length** We assume that a population occupies a circular area, so its diameter ($d_{po}$)

can be derived from the population area ($S_{po}$) as:

$$d_{po} = 2 \times \sqrt[2]{\frac{S_{po}}{\pi}} \tag{S26}$$

where π is set to 3.14.

**Volume** The population volume ($V_{po}$) is calculated as area multiplied by thickness. The area is assumed to be circle, and the thickness is set to 100 m above the ground surface:

$$V_{po} = S_{po} \times (100) \tag{S27}$$

where π is set to 3.14.

**Mass** The population mass ($m_{po}$) is down-scaled from the national total mass according to the ratio of the population surface area ($S_{po}$) to the ecosystem surface area ($S_{es}$).

$$m_{po} = \frac{S_{po}}{\pi \times {r_{es}}^2} \times m_{es} \tag{S28}$$

where π is set to 3.14, $r_{es}$ is the radius of the ecosystem, and $m_{es}$ is the mass of the ecosystem.

**(11) Ecosystem**

Ecosystem refers to the unified whole formed by interactions and interdependencies between biotic and abiotic components within a given time and space through continuous material flow, energy flow and information flow, which constitutes an ecological unit of structure, function and evolution (Pavé, 2006; Chang and Ge, 2010).

**Length** The length of the ecosystem ($d_{es}$) adopts the estimate for aquatic ecosystems reported by Post et al. (2000) and Post (2002), with an average value of 18,800 m.

**Volume** Ecosystem volume ($V_{es}$) is calculated as area multiplied by thickness. The area is assumed to be circular, and the thickness is set to 100 m above the ground surface and 32 m below the ground surface (Post, 2002):

$$V_{es} = \pi \times {r_{es}}^2 \times (32 + 100) \tag{S29}$$

where π is set to 3.14.

**Mass** The ecosystem mass ($m_{es}$) represents the total mass of water bodies and aquatic organisms. Given that organisms account for a small fraction of lake water mass, water mass is used as a substitute:

$$m_{es} = \pi \times {r_{es}}^2 \times (32 \times \rho_{water} + 100 \times \rho_{air}) \quad \text{(S30)}$$

where π is set to 3.14; $r_{es}$ is the radius of the ecosystem; $\rho_{water}$ denotes the water density, which is set to $10^3$ kg/m$^3$.

**(12) Organara of eukarcity**

Various facilities that supply goods and services for humans in cities are defined as city functional components (Xu et al., 2020). Among them, closed systems with high material and energy fluxes are termed organaras (Chang et al., 2021). Referring to de Groot et al. (2002), Xu et al. (2020) divided complete ecosystems into five functional groups and classified organaras into six categories: regulator, circulator-transporter, life maintainer, primary producer, secondary producer, and decomposer. In this study, 16 types of organaras with a total number of 6,170 were collected (Yang et al., 2021; Chang et al., 2022; Wu et al., 2023; Zhou et al., 2025; our unpublished data), including: circulator-transporter: 490 shopping malls, 88 express delivery outlets, 75 gas stations; life maintainer: 22 swimming pools, 7 gymnasiums, 7 convenience stores, 437 catering restaurants, 666 agritourism ecosystems; regulator: 131 banks, 65 data centers; primary producer: 3,546 greenhouses, 24 vertical farms; secondary producer: 490 dairy farms, 15 thermal power plants; decomposer: 62 wastewater treatment plants, 45 constructed wetlands (Table S1).

**Area** In this study, the area of organaras ($S_{oa}$) was obtained through field surveys and spatial identification based on Google Earth, and the mean value was adopted.

$$S_{oa} = \frac{\sum_{i=1}^{N} S_{oai}}{N} \quad \text{(S31)}$$

where $S_{oai}$ denotes the area of the $i$-th organaras, and $N$ is sample size.

**Length** We conceptualize each organaras as a cube, and its length $d_{oa}$ is defined as:

$$d_{oa} = 2 \times \sqrt[2]{S_{oa}} \quad \text{(S32)}$$

**Volume** The organaras volume $V_{oa}$ is calculated as area multiplied by height:

$$V_{oa} = S_{oa} \times H_{oa} \quad \text{(S33)}$$

where $H_{oa}$ is the height of the organara, which is set to 4.87 m (Xu et al., 2020).

**Mass** In this study, the organaras mass ($m_{oa}$) was derived from the mean mass values of circulation-type express delivery outlets, regulation-function data centers, and secondary-processing dairy farms. The mass of an organaras is calculated as volume multiplied by density (Wu et al., 2025). The density of buildings was obtained from the Building Construction Handbook (6th edition).

$$m_{oa} = \frac{\sum_{i=1}^{N} m_{oai}}{N} \tag{S34}$$

where $m_{oai}$ denotes the mass of the *i-th* organaras, and $N$ is sample size.

**(13) Eukarcity**

The eukarcity refers to the coupled urban-rural system, which analogous to eukaryotic cells, the EC consists of citynucleus, organaras, and cityplasm (Chang et al., 2021). Its horizontal boundary corresponds to the administrative boundary in densely populated regions (Chang and Ge, 2005), because policy formulation and implementation are consistent within administrative divisions and differ from those of other cities (Montgomery, 2008; Chang et al., 2022). Markers for EC boundaries are control points such as checkpoints on major transportation routes and toll stations (Yang et al., 2015). The upper vertical boundary of the EC is set at 1,000 m above the ground surface, while the lower boundary is the rock-stratum surface, the deepest ocean depth, and the maximum drilling depth, which is set to 10,000 m. Considering demographic characteristics, prefecture-level cities in China are comparable to the Metropolitan Statistical Areas (MSA) in the United States and Metropolitan Areas (MA) in Europe (Zünd and Bettencourt, 2019).

**Area** The area of a eukarcity ($S_{ec}$) is the administrative-unit area of coupled urban-rural system, given that policy formulation and implementation are consistent within administrative boundaries. The sample of cities are collected from China ($N$ = 277), Metropolitan Statistical Areas (MSAs) in the United States ($N$ = 384), and Metropolitan Areas (MAs) across five European countries (France, $N$ = 15; United Kingdom, $N$ = 14; Germany, $N$ = 24; Spain, $N$ = 8; Italy, $N$ = 11), yielding a total sample of $N$ = 733 (NBSC, 2019; U.S. Census Report, 2022; Bettencourt and Lobo, 2016). The average eukarcity area ($S_{ec}$) is

$$S_{ec} = \frac{\sum_{i=1}^{N} S_{eci}}{N} \tag{S35}$$

where $S_{eci}$ denotes the administrative-unit area of prefecture-level cities, and $N$ is sample size.

**Length** We conceptualize each eukarcity as a circle, and its length ($d_{ec}$) is

$$d_{ec} = 2 \times \sqrt[2]{\frac{S_{ec}}{\pi}} \tag{S36}$$

where π is set to 3.14.

**Volume** The eukarcity is conceptualized as a cylinder. In terms of thickness, it covers 1,000 m above the ground surface and 10,000 m below the ground surface (Baker et al., 2001; Gu et al., 2009; Jamieson et al., 2010). The calculation formula for eukarcity volume $V_{ec}$ is:

$$V_{ec} = S_{ec} \times 11000 \tag{S37}$$

**Mass** The eukarcity mass ($m_{ec}$) is taken from Wu et al. (2025), the mean total mass (wet biomass mass plus anthropogenic mass) of Chinese cities,

$$m_{ec} = \frac{\sum_{i=1}^{N} m_{eci}}{N} \tag{S38}$$

where $m_{eci}$ denotes the mass of the *i-th* eukarcity, and $N$ is the sample size.

**(14) Biosphere**

The biosphere is the entirety of all living organisms on Earth, referring to the sphere composed of life forms inhabiting the surface of the lithosphere, the lower part of the atmosphere, the hydrosphere, and the pedosphere (Suess, 1875; Odum and Odum, 1959; Gillard, 1969; Chang and Ge, 2010).

**Area** The biosphere area ($S_{bs}$) is set equal to the surface area of the Earth, which is about $5.1 \times 10^{14}$ $m^2$ (National Geospatial-Intelligence Agency, 2014).

**Length** We consider the biosphere as a circular plane, and its length is the diameter of the circle ($d_{bs}$),

$$d_{bs} = 2 \times \sqrt[2]{\frac{S_{bs}}{\pi}} \tag{S39}$$

where $S_{bs}$ denotes the surface area of the Earth, and π is approximated as 3.14.

**Volume** The biosphere is not only the region where life exists, but also the core of geochemical processes. It encompasses the pedosphere and the hydrosphere, and is distributed across the upper layer of the lithosphere and the lower layer of the atmosphere. Consequently, its upper boundary is set at $10^4$ m above the Earth's surface, and its lower boundary at $10^4$ m below the surface (Chang and Ge, 2010; Jamieson et al., 2010). In calculations, the biosphere can be regarded as a spherical shell enclosed by two concentric spheres, and its volume is obtained by subtracting the volume of the inner sphere from that of the outer sphere,

$$BSV_{bs} = \frac{4}{3} \times \pi \times ((R + 10^4)^3 - (R - 10^4)^3) \tag{S40}$$

where $R_E$ denotes the Earth's mean radius.

**Mass** The biosphere mass $m_{bs}$ is defined as the wet biomass of all organisms on Earth, which is ~2.2 × $10^{18}$ g (Elhacham et al., 2020).

## Text S2 Data collection for energy metabolism of living systems

The energy metabolisms of prokaryotes, unicellular eukaryotes, multicellular organism (compiled by Wu et al., 2025), and eukarcities have been estimated (Wu et al., 2025). We calculate the energy metabolism of the Earth as the total power of solar radiation received by the Earth,

$$P_{AE} = I_{sun} \times A_E \tag{S41}$$

where, $P_{AE}$ is the energy metabolism of the anthro-Earth, $I_{sun}$ is the standard total solar irradiance (the solar constant), which is $1{,}361\ \mathrm{W} \cdot m^2$ (Prša et al., 2016), $A_E$ is the hemisphere facing the Sun, which is $1.275 \times 10^{14}$ $m^2$ (Archer, 2011).

## Text S3 Selection and standardization of the traits of the 14 known levels of living systems

We chose the 14 levels of living systems, including biomacromolecule subunit, biomacromolecule, biomacromolecule complex, prokaryotic cell, organelle, eukaryotic cell, tissue, organ, multicellular organisms, population, organara, ecosystem, eukarcity, and biosphere. We collect the characteristic attributes shared by each level across three dimensions: structure (including morphology), process, and function. The selection, scoring, and formal definition of these systemic traits were grounded in the Phenotype And Trait Ontology (PATO; Gkoutos et al., 2018; http://purl.obolibrary.org/obo/pato) and the Systems Biology Ontology (SBO; Courtot et al., 2011; https://www.ebi.ac.uk/sbo/), ensuring cross-scale data standardization and robust mathematical comparability across drastically different dimensions (Table S2).

To accommodate the highly diverse biological and structural attributes across the entire living spectrum, specific traits were rigorously cross-referenced and mapped to established open-source domain ontologies. Biomolecular subunits, biomolecules, and biomolecule complexes reused the Chemical Entities of Biological Interest (ChEBI; https://www.ebi.ac.uk/ols4/ontologies/chebi, Hastings et al., 2016) and the Gene Ontology (GO; The Gene Ontology Consortium, 2019; https://www.ebi.ac.uk/ols4/ontologies/go); organelles, prokaryotic cells, and eukaryotic cells reused the GO and Ontology of Microbial Phenotypes (OMP; https://www.ebi.ac.uk/ols4/ontologies/omp) and the Cell Ontology (CL, https://www.ebi.ac.uk/ols4/ontologies/cl); attributes of tissues, organs, and organisms referred to the Uber-anatomy Ontology (UBERON, https://www.ebi.ac.uk/ols4/ontologies/uberon), the NCBI Organismal Taxonomy (NCBITaxon, https://www.ebi.ac.uk/ols4/ontologies/ncbitaxon), and the Plant Ontology (PO, https://www.ebi.ac.uk/ols4/ontologies/po); populations and ecosystems referred to the Population and Community Ontology (PCO, https://github.com/PopulationAndCommunityOntology/pco) and the Environment Ontology (ENVO, https://www.ebi.ac.uk/ols4/ontologies/envo). Meanwhile, by borrowing the hierarchical construction principles of organelle and cell ontologies, we specifically developed the Organara Ontology (OAO), the Eukarcities Ontology (ECIO) and Biosphere Ontology (BSO), detail see online supplementary material, https://github.com/smanalysis/Ontology, thereby achieving the standardized quantification of the organaras and city levels within the same comparative matrix. In total, 16 traits were identified, comprising 3 quantitative traits and 13 qualitative traits (Table S3, S4). Quantitative traits were measured as specific numerical values or log-transformed values ($\log_{10}$); qualitative traits were assigned scores on binary or ordinal scales following the above ontology.

**Text S4 Defining levels of living systems**

The characteristics of the hierarchical living system are multi-level, scale from small to large, nested, self-similarity, diverse, complex, and identical. The living systems at different levels have specific temporal and spatial scales. After the 4 levels of living systems: gene, organism, population, and ecosystem were drawn by Odum and Odum (1959), many studies have expanded the spectrum of living system (Chang and Ge, 2001; Pavé, 2006; Jørgensen et al., 2016; West, 2017; Donges et al., 2021). However, some levels of these systems have no clear boundaries. Here we list some "system" for example, a landscape is a complex of various ecosystems and it is a research unit but not a living system, and landscape ecology studies the structure and dynamics across any number of multiple ecosystems (Chang and Ge, 2005). A biome is a great biological community within a clime region (Roxburgh and Noble, 2001; Mucina, 2019) and it has no clear boundary distinct with other biomes.

In present understanding, some levels of living systems are confusing in size. For example, a biological community has space with no function while an ecosystem only has function with no space, so a biological community is the hardware of an ecosystem (Chang and Ge, 2003, 2005, 2010). How to identify a living system that distinct with others is an important question. In 1944, Schrodinger used to refine a 'living system' in respect of the information entropy in thermodynamics in the famous book "What is Life" (Moberg, 2020). Prigogine proposed the dissipative structure theory in the 1960s for self-organizing systems that exchange energy and materials with their environment (Prigogine, 1967). Chang and Ge (2001, 2005) proposed strict criteria in structure, process and function of a living system: (1) clear boundary, within which the structure and processes are distinct, (2) three dimensions of space and one dimension of history, (3) self-organization, which is the processes within the system and resist external disturbances (4) information transmission and feedback with minimal delay.

**Text S5 Allometric relationships between the quantitative traits of living systems**

In order to find the scale laws between the traits, we analysis the allometric relationship among the levels in the quantitative traits,

$$Y = \alpha x^{\beta} \quad (S42)$$

where, $\alpha$ is the coefficient, $\beta$ is the scaling power.

All the correlations for length, mass, volume and time of the living systems are significant ($p < 0.05$) except biosphere (Fig. S1 a-f). The mass ($m$) in response to length ($l$) is $m \propto l^{3.37}$ from BMS to eukaryotic cell, while, that for multi-cellular tissues and higher-level living systems decrease to $\beta$~2.37. That is, scaling law for mass to length increase is highly superliner. The break point between unicellular and multicellular tissue show they follow distinct scaling law. An overall model for all the 14-level is exist ($p < 0.05$). The mass in response to volume also follows two scaling models for unicellular and below-level systems, and multicellular and above system, respectively. The volume in response to length of living systems almost no break, in other words, 14 levels follow one model. On the contrary, the active duration $t$ in response to volume are from two distinct models (Fig. S1 d), $t \propto v^{0.71}$ for unicellular and below levels and $t \propto v^{0.20}$ for multi-cellular tissues and the above level systems. Similarly, the active duration t in response to length l is also best described by two distinct scaling models, $t \propto l^{2.11}$ for unicellular and below-level systems and $t \propto l^{0.64}$ for multicellular and above system (Fig. S1 e). The active duration in response to mass shows the same pattern, with $t \propto m^{0.57}$ for unicellular and below-level systems and $t \propto m^{0.26}$ for multicellular and above-level systems (Fig. S1 f). The allometric power for unicellular living systems is higher than that of multicellular living systems, and it is similar to the energy metabolism in response to mass is high in unicellular, and getting lower in multicellular and mammals (DeLong et al., 2010). Meanwhile, biosphere does not fit the correlations of other levels of living systems, and it means the allometric model end in eukarcity. It reveals that the scale laws are distinct from unicellular to multicellular living systems in evolution.

**Text S6 Theoretical syntheses of some existing models and principles by the HELIX**

HELIX is compatible with many present theories, principles and models that for a few cross-level living systems. Here we discuss some major theories and models for cross-level of living systems that synthesized by the HELIX.

***Homeostasis*** It is the mechanism of self-regulating steady state of physical and chemical factors of a living system when face environmental changes (Cannon, 1929; Davies, 2016). The typical one is the temperature homeostasis of warm-blooded animal (mammals and birds), as well as water and salt homeostasis of cells and organisms (Chang and Ge, 2001; Terrien et al., 2011; Rowland, 2002). The relative stable statuses ensure the biological-chemical reactions and enhance the resistance to disturbances. Similar to organism, modern cities in industrialization are found to have been developed to eukarcities (Chang et al., 2021). In context of the HELIX, the living systems in group autonomous all have many factors of homeostasis. For example, the light homeostasis is modern cities keep suitable light intensity in night so keep a stable light in 24 hours a day. Other examples are the stable tap water in anytime, stable food and power supply in urban areas and rural areas. Notably, the 'heat island effect' in urban areas can be seen as the temperature homeostasis, which similar to warm-blooded animals, of modern cities for it minimizes the temperature differences of urban among the four seasons, as well as day and night within urban areas (Chang et al., 2021). Besides, the homogenizing effect among the urban green spaces also performs in the convergence of cities in different climate regions (Chen et al., 2021; Yang et al., 2023). In theory, the HELIX synthesize all period of the autonomous living systems have the homeostasis due to the self-organization.

***GAIA hypotheses*** The GAIA means the co-reactions of the five spheres balance and provide stable physical and chemical conditions that suitable for livings (Margulis and Lovelock, 1989; Lovelock, 1990). Biosphere is the central controller among geosphere, atmospheric, pedosphere and hydrosphere (Lovelock and Margulis, 1974). In context of the HELIX, GAIA is a control-feedback mechanism that the Earth surface maintaining homeostasis in temperature, atmospheric oxygen levels, ocean salinity, and other factors of the five major spheres (Chang and Ge, 2005). A further 'Daisy World Model' of GAIA is trying to explain the self-regulation of the homeostasis in most extent (Watson and Lovelock, 1983). However, GAIA has repeatedly failed to maintain stability, resulting in five mass extinction events (Barnosky, 2011). In context of the HELIX, GAIA is the self-organization mechanism for the homeostasis of the Earth but it is not a 'super organism'.

***Meme*** The concept 'Meme' in promoted in the famous book "The Selfish Gene" (Dawkins, 1976). A memes is a unit of human culture, including science, engineering, production, life style, literature and art, as well as oral presentation (Blackmore, 2000; Mesoudi et al., 2006; Mesoudi, 2017). The meme refer to various human cultural factors in the anthroposociology and has widely cited and already embodied in the Oxford English Dictionary after it appeared for only 30 years. However, there are increasingly arguments and it developing to the 'informal' culture recently. For example, a meme have neither material substance corresponding to a gene, nor the specific transmission approach and duplication mode in genetics (Claidière and André, 2012; Gray and Watts, 2017; Ermakov and Ermakov, 2021) so it is hardly keep the fidelity in informational transferring. In context of the HELIX, a meme is a mixed concept including the formal coding information (similar to the gene stored in DNA) and the informal information (similar to the RNA). The recent definition of meme is 'a sharable cultural template that proliferates on the internet' (Shabunina and Pasi, 2018; Galip, 2024). The internet meme usually uses digital combine visual media, text, audio, action, gesture and narration in a humorous way. Therefore, it is necessary to perform functional decomposition on memes to distinguish different types of human cultural information units.

***Ecosystem services*** The goods and services obtained from ecosystems by humans are the basis of human well-being (Costanza et al., 1997; Cumming et al., 2014; Chang et al., 2021; Daily and Ruckelshaus, 2022). Ecosystem services are an account of the market value of the functioning of ecosystems (Chang et al., 2011; Costanza et al., 2014). The commonly used categories of ecosystem services follow the Millennium Ecosystem Assessment framework, namely provisioning, regulating, and cultural ecosystem services (MA, 2005). Artificial (or engineering) ecosystems refer to the ecosystems that have been transformed from natural ecosystems, ranging from slightly disturbed to completely rebuild by humans (Palmer et al., 2004). Chang et al. (2021) divide the ecosystem services provided by artificial ecosystems into target ecosystem services and accompanying ecosystem services. The target services result from investment for specific economic goals. Some ecosystem services outside the investment goal, such as biomass production outside the grain production, are the accompany ecosystem services (Chang et al., 2021, 2022), which are equivalent to externalities in economics. The sub-categories of target and accompanying ecosystem services follow the Millennium Ecosystem Assessment framework (MA, 2005). The target ecosystem services are calculated based on the revenue of an artificial ecosystem (Chang et al., 2017; Wu et al., 2024), while the calculation of accompanying services is the same as that of natural ecosystems.

In the context of the HELIX, the 'services' link the two groups CS and AS. That is, the living systems in CS provide 'services' to their mother living systems (in other words, higher level system) in AS. For example, ecosystems, either natural or artificial, provide goods and services for humanity (Palmer et al., 2004; Cumming et al., 2014) in eukarcities (Chang et al., 2021). Organara is the enhanced artificial ecosystem for providing ecosystem services driven by human investment (Chang et al., 2021, 2022). Due to the decline of natural ecosystems and their services (Raudsepp-Hearne et al., 2010), most ecosystem services are provided by organaras in urban areas so the Organara-Based Management (OBM) has become the major loop between human and ecosystems (Chang et al., 2021). OBM coupled with the Ecosystem-Based Management (EBM) loop in the anthropogenic world. Similar to ecosystem services, all levels of living systems of CS provide 'biological services' to their mother living systems in AS.

***Endosymbiosis*** The theory promotes that eukaryotic cells are originated from the eubacteria engulfing anaerobic microorganisms (Margulis, 1975, 1981). For example, within a big 'cell', the Engulfment of cyanobacteria form chloroplasts while spirilla to be mitochondria, they also keep their own gene copying beside the eukaryote core --- semi-autonomous copy of genes. Margulis (2010) further proposed the fusion to the formation of prokaryotic cells as the 1st fusion, then the formation of eukaryotic cell as the 2nd fusion, and multicellular organisms, especially gut microbiota, as the 3rd fusion (Table S9). In sum, in the context of the HELIX, the three fusions in Endosymbiosis can all be explain as the origin mechanisms of the autonomous living systems.

The HELIX extends the new emerged eukarcity as the 4th fusion the organaras and citynucleus and cityplasm to be a living system (Chang et al., 2021). Similar to the organelles in eukaryotic cell, some organaras also possess and copy their own technology and other information beside the information from citynucleus so the semi-autonomous replicate also exists. For example, manufactories, schools and banks, as well as the enclosed dairy farms, carry their own information and self-replication beside the information from citynucleus (Chang et al., 2021). The concept of eukarcity clarifies several aspects in evolution, ecology and geology.

## Text S7 Data collection and calculation for the size of some newly emerged living systems

### (1) Anthropogenic Earth Living System

**Length** The length of anthro-Earth is the diameter ($d_{AE}$), defined as twice the distance from the Earth's center to the outer boundary of the artificial satellite sphere.

**Volume** the anthro-Earth can be viewed as a spherical shell enclosed by two spherical surfaces and its volume ($V_{AE}$) can be obtained by subtracting the inner sphere from the outer sphere,

$$V_{AE} = \frac{4}{3} \times \pi \times \left(\left(R_{big}\right)^3 - (R_{small})^3\right) \quad \text{(S43)}$$

where $R_{big}$ is the radius of the outer sphere, determined by the upper boundary, $6.4 \times 10^6 + 10^7 = 1.67\times10^7$ m; and $R_{small}$ is the radius of the inner sphere, determined by the lower boundary, $6.4 \times 10^6 - 1 \times 10^4 = 6400{,}000 - 10{,}000 = 6{,}390{,}000$ m, approximately $6 \times 10^6$ m.

**Mass** The mass ($m_{AE}$) is the sum of the masses of the biosphere, hydrosphere, pedosphere, and anthropogenic mass,

$$m_{AE} = m_{bs} \times 2 + m_{water} + m_{soil} \tag{S44}$$

where $m_{bs}$ is the mass of the biosphere, $m_{water}$ is the mass of the hydrosphere, and $m_{soil}$ is the mass of the pedosphere. The anthropogenic mass on the Earth surpassed the total global biomass, which is approximately $2.2 \times 10^{18}$ g, around the year 2020 (Elhacham et al., 2020).

**(2) Country**

**Length** The country is set as a circular plane, with its length ($d_{ct}$),

$$d_{ct} = 2 \times \sqrt[2]{\frac{S_{ct}}{\pi}} \tag{S45}$$

where $S_{ct}$ is the national land areas and π is taken as 3.14.

**Volume** We treat the country as a cylinder, whose thickness is set equal to the biosphere thickness, which are $10^4$ m for both above and below the ground surface. Hence, the volume ($V_{ct}$) is,

$$V_{ct} = S_{ct} \times 2 \times 10^4 \tag{S46}$$

**Mass** The mass of a country ($m_{ct}$, wet weight) is defined as the sum of the biomass mass and the anthropogenic mass. This mass is then downscaled from the total Anthropogenic Earth Living System mass $m_{AE}$ according to the ratio of the national land area $S_{ct}$ to the Earth's land surface area,

$$m_{ct} = \frac{S_{ct}}{S_{bs} \times 0.29} \times m_{AE} \tag{S47}$$

where $S_{bs}$ is the Earth's surface area, and 0.29 is the fraction of the land area relative to the total surface area.

The areas ($S_{ct}$) of country vary considerably, with the vast majority of countries having relatively small areas. To mitigate the influence of the long-tail effect (Apergis and Payne, 2012), we selected the top 60 countries by area (Apergis and Payne, 2012; Bhattacharya et al., 2016) based on data published by the World Bank in 2022.

**(3) City cluster**

The city cluster (urban agglomeration) is a spatially contiguous group of eukarcities that are linked by intensive flows of people, goods, capital and information, and that are planned and governed as one functional region (Gottmann, 1961; Dash Nelson and Rae, 2016; Fang and Yu, 2017).

**Length** We treat the city cluster area as a circle, and its length ($d_{cc}$) is,

$$d_{cc} = 2 \times \sqrt[2]{\frac{S_{cc}}{\pi}} \tag{S48}$$

where π is taken as 3.14.

**Volume** The city cluster is modeled as a cylinder. In the vertical direction, following the urban standard (Gu et al., 2009), the upper boundary is set at 1,000 m above ground level, and the lower boundary is set at the rock surface, the deepest ocean depth, and the deepest drilling depth, taken as $10^4$ m (Gu et al., 2009; Jamieson et al., 2010). Therefore, its volume ($V_{cc}$) is the area multiplied by the thickness (the sum of the upper and lower boundaries),

$$V_{cc} = S_{cc} \times 11000 \tag{S49}$$

**Mass** The mass of the city cluster ($m_{cc}$) is,

$$m_{cc} = \frac{S_{cc}}{S_{ec}} \times m_{ec} \tag{S50}$$

where $S_{ec}$ is the mean eukarcity area and $m_{ec}$ is the mean eukarcity mass.

The area of a city cluster ($S_{cc}$) is the total area of its member administrative units, taken from official planning documents and national statistical delimitations.

$$S_{cc} = \frac{\sum_{i=1}^{N} S_{cci}}{N} \tag{S51}$$

where $S_{cci}$ is the area of the $i$-th city cluster, and N is sample size.

The sample includes 13 city clusters. China (N = 4): the Yangtze River Delta (26 cities; NDRC and MOHURD, 2016), the Guangdong–Hong Kong–Macao Greater Bay Area (CPC Central Committee and State Council, 2019), Beijing–Tianjin–Hebei (NDRC, 2019) and the Beibu Gulf (NDRC and MOHURD, 2017) city clusters. United States (N = 2): the Northeast megaregion (Regional Plan Association, 2009) and the New York–Newark Combined Statistical Area (U.S. Census Bureau, 2023; Wikipedia, 2026d). Other countries (N = 7): the National Capital Region of Japan (Statistics Bureau of Japan, 2010); the Seoul Capital Area (NGII, n.d.); Frankfurt Rhine-Main (Wikipedia, 2026a); the Moscow agglomeration, comprising the city of Moscow and Moscow Oblast (Wikipedia, 2026c); the Greater Golden Horseshoe (Wikipedia, 2026b); the Mexico City megalopolis (INEGI, 2020); and the São Paulo Macrometropolis (São Paulo State Government, 2017). Where no official area figure was available, the area was taken from the corresponding Wikipedia entry.

**Text S8 Artificial infrastructure, energy and information of the anthro-Earth**

AE shows the traits of AS group in three aspects. (1) Along with humanity enriching the mass and chemical composition of the Earth's living systems, the Earth has also continuously added new substances. Starting from the eukarcity, humans have expanded the major chemical elements of living systems. For example, artificial infrastructure including houses and roads made from materials such as cement, glass, and ceramics are mainly composed of silicates; the fiber-optic communication equipment and photovoltaic power generation are also derived from silicon. Crucially, semiconductor materials like silicon have entered the realm of digital and intelligent information regulation. Living systems have transitioned from the 'carbon-based system' (Lovelock, 2003; Rampelotto, 2010) to the 'carbon-silicon composite' system (Chang and Ge, 2005). This marks a significant milestone in the evolutionary process of life.

(2) A continuous increase of about 30% in solar radiation over time in past 4.6 billion years, and a great increase of energy input to the Earth's surface (Sagan and Mullen, 1972; Bahcall et al., 2001). However, the Earth maintains a moderate state of stability regulated by the GAIA. Furthermore, Humanity has release great amount energy by of fossil fuel since the Industrial Revolution in the past 300 years. People now transport energy (oil, natural gas, electricity, and coal) over thousands of kilometers (Chang et al., 2021). Humanity has changed the global biogeochemical cycles and energy flow by trade profoundly. The material flux driven by Humanity on the Earth is 2-4 orders of magnitude higher than that driven by natural force (Gu et al., 2013; Chang et al., 2021). Personnel flows are becoming increasingly global (Niva et al., 2023). Nowadays, the material and energy fluxes have been transformed from biogeochemical cycle to the 'Anthropo-biogeochemical cycle' (Chang and Ge, 2005, 2010). Although ecological problems and environmental pollutions are exiting, human efforts try to keep stability of the Earth. Humanity reinforced the biological regulatory GAIA to upgrade to an 'Anthropogenic GAIA', which we define as 'AGAIA' in this paper. AGAIA is the control-feedback mechanism for coupling biosphere and other spheres and driven by both natural and human cultural forces since industrialization. Beside the positive significance, AGAIA is facing many challenges, such as human population explosion, climate change, resource overexploitation and socioeconomic fluctuations, as well as biosphere integrity, biogeochemical flows, freshwater change and land-system change (Steffen et al., 2015). Many presses are thought exceeding the safe limits of the Earth (Richardson et al., 2023). Local improvements at the global scale may still constitute an unsustainable trap with largely unrepaid ecological debts (Arrow et al., 1995; Cumming et al., 2014). The human dominated organara based management (Chang et al., 2021) will improve the ecology condition of the anthro-Earth when coupling with ecosystem based management.

(3) Human cultural information, which can be divided as Meme (Dawkins, 1976; Blackmore, 2000), and more precisely cene (token) and tene (Chang and Ge, 2005), has become the major regulator of the anthro-Earth. Cene and tene regulates persons, organizations, cities and countries by two channels. First, electronic telecommunication is start from telegram in the 1930s by Morse code. After that, telephone, fax and e-mail to be the major and now instant messaging are the major. The electronic communication is similar to the nerves, which is also transforming by electronics. Second, human made energy flow and logistics, as well as monetary flow regulate the world. This channel is similar to the material flow and hormones of animals. The delay of global communication in the 4$^{th}$ generation mobile communication technology (4G) is 20~50 ms ($10^{-2}$ s) with the technology within two decades, while the delay <1 ms ($10^{-3}$ s) in 5$^{th}$ generation mobile network (5G) within one decade to now. The satellite-ground communicate is 1-270 ms in present. The delay-free of telecommunication ensure human culture has been the core regulator on the anthro-Earth living system.

## Supplementary figures

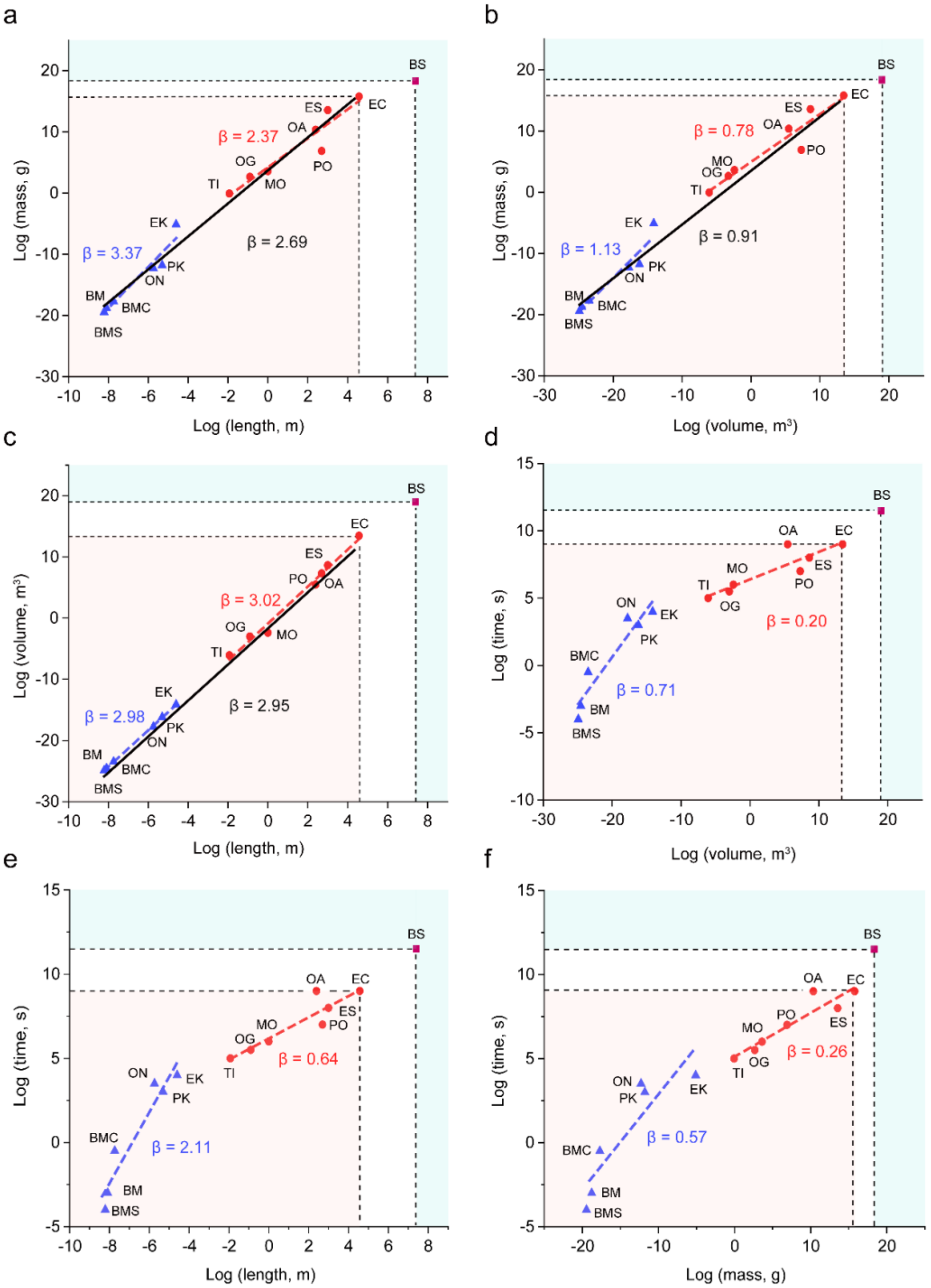

**Fig. S1 Allometric relationship between the four quantitative traits of 14 levels of living systems.** (a) mass (wet weight) and length, (b) mass and volume, (c) volume and length, (d) time and volume, (e) time and length, (f) time and mass. Solid line denotes fitted model across the levels below the biosphere, dashed lines indicate piecewise models. All the fitting models are significant ($p < 0.05$).

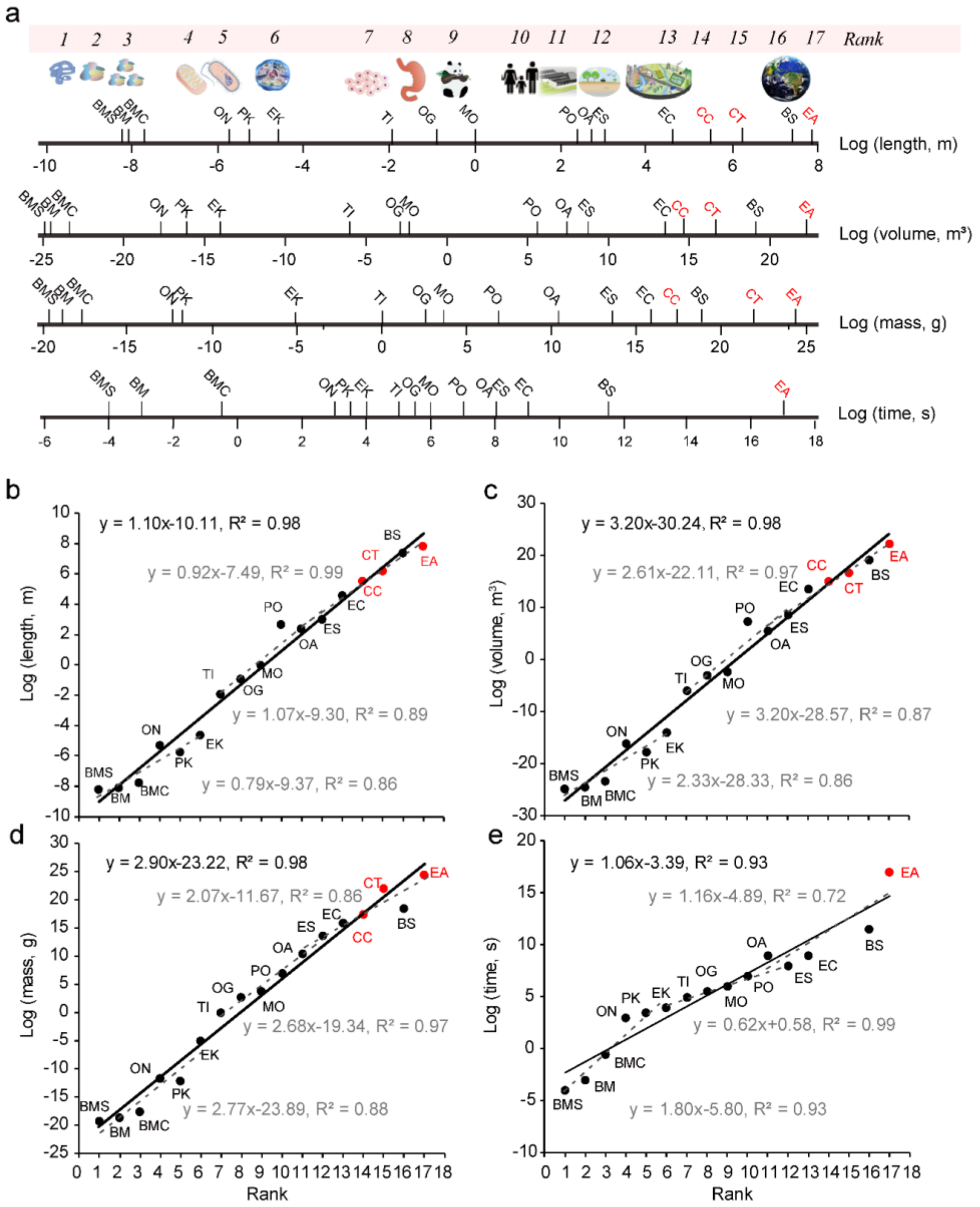

**Figure S2 Quantitative traits of the 17 levels of hierarchical living systems and their exponential models with level rank.** (a) Position of the 17 levels on length, mass, volume and time, numbers in the top row give the hierarchical rank, (b-e) relationships between the rank and length (b), volume (c), mass (d), and time (e), solid lines indicate models fitted across all levels, and grey dashed lines show the piecewise models. Abbreviations and rank: 1 BMS, biomacromolecule subunit, 2 BM, biomacromolecule, 3 BMC, biomacromolecular complex, 4 PK, prokaryotic cell, 5 ON, organelle, 6 EK, eukaryotic cell, 7 TI, tissue, 8 OG, organ, 9 MO, multicellular organism, 10 PO, population of MO, 11 OA, organara, 12 ES, ecosystem, 13 EC, eukarcity, 14 CC, city cluster, 15 CT, country, 16 BS, biosphere, 17 AE, anthro-Earth. All the models are significant ($p < 0.05$).

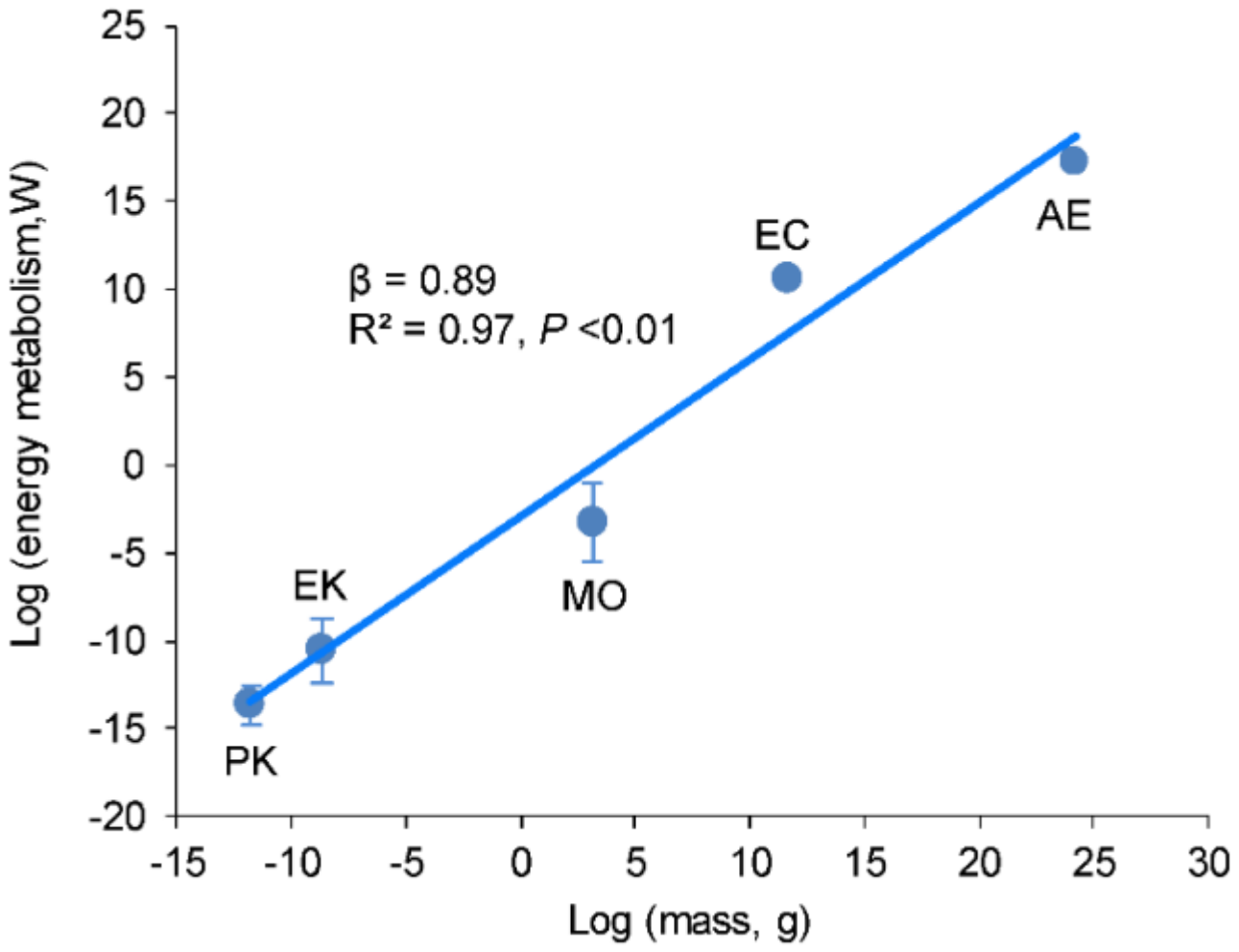

**Fig. S3 Scale of the energy metabolism in response to the mass (wet weight) of the living systems in AS.**

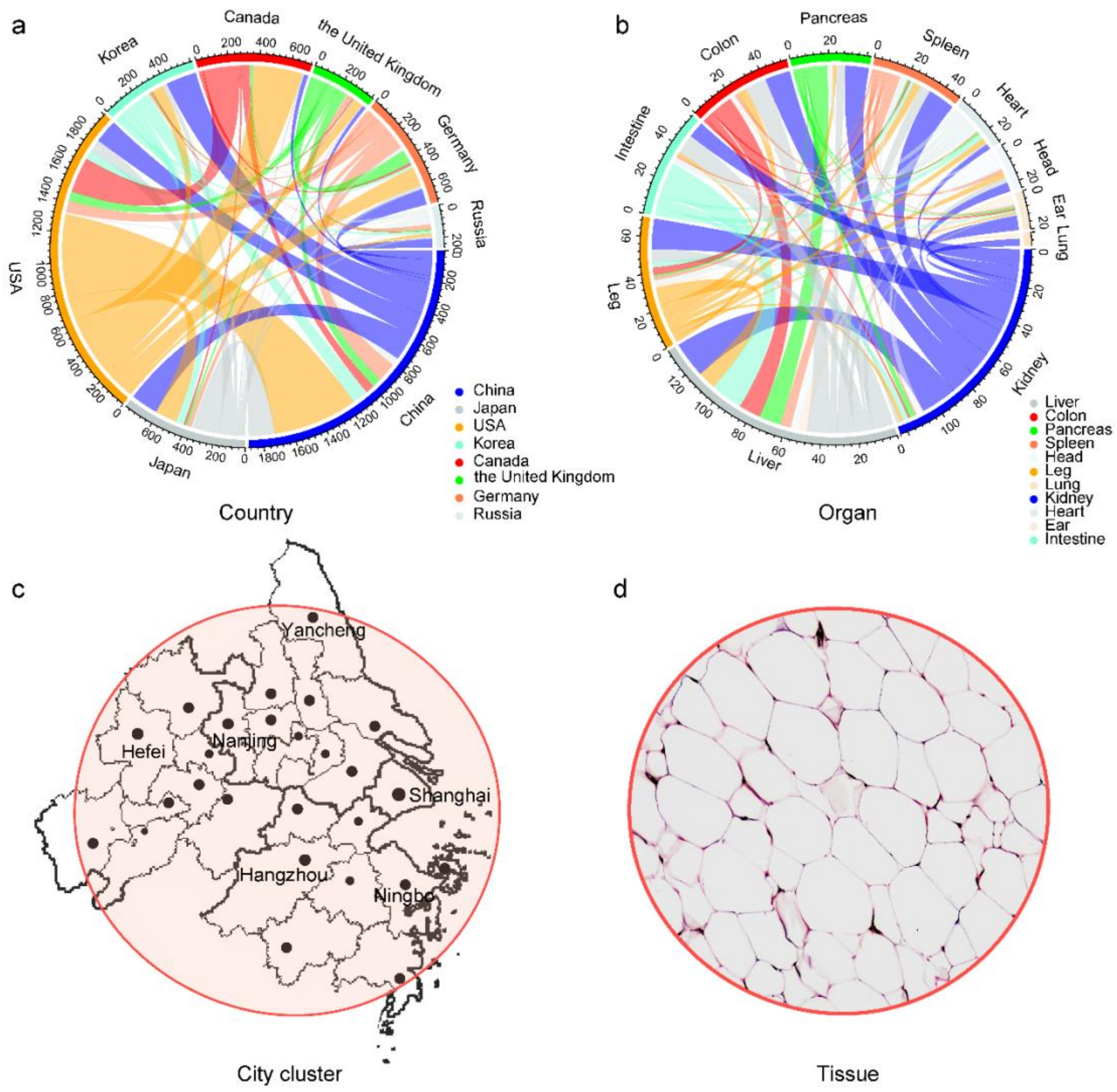

**Fig. S4 Trans-level comparisons of the new levels of living systems within their groups.** (a-b) interactions between the components, (a) trades between countries and (b) material flows between organs; (c-d) morphological similarities between city cluster and tissue. (c) Yangtze River Delta City Cluster, China, irregular line represents the administrative boundary of eukarcities, and the shaded circle area simulates the spherical structure of the eukarcity, (d) cross-section of adipose tissue (image courtesy of Berkshire Community College Bioscience Image Library, CC0).

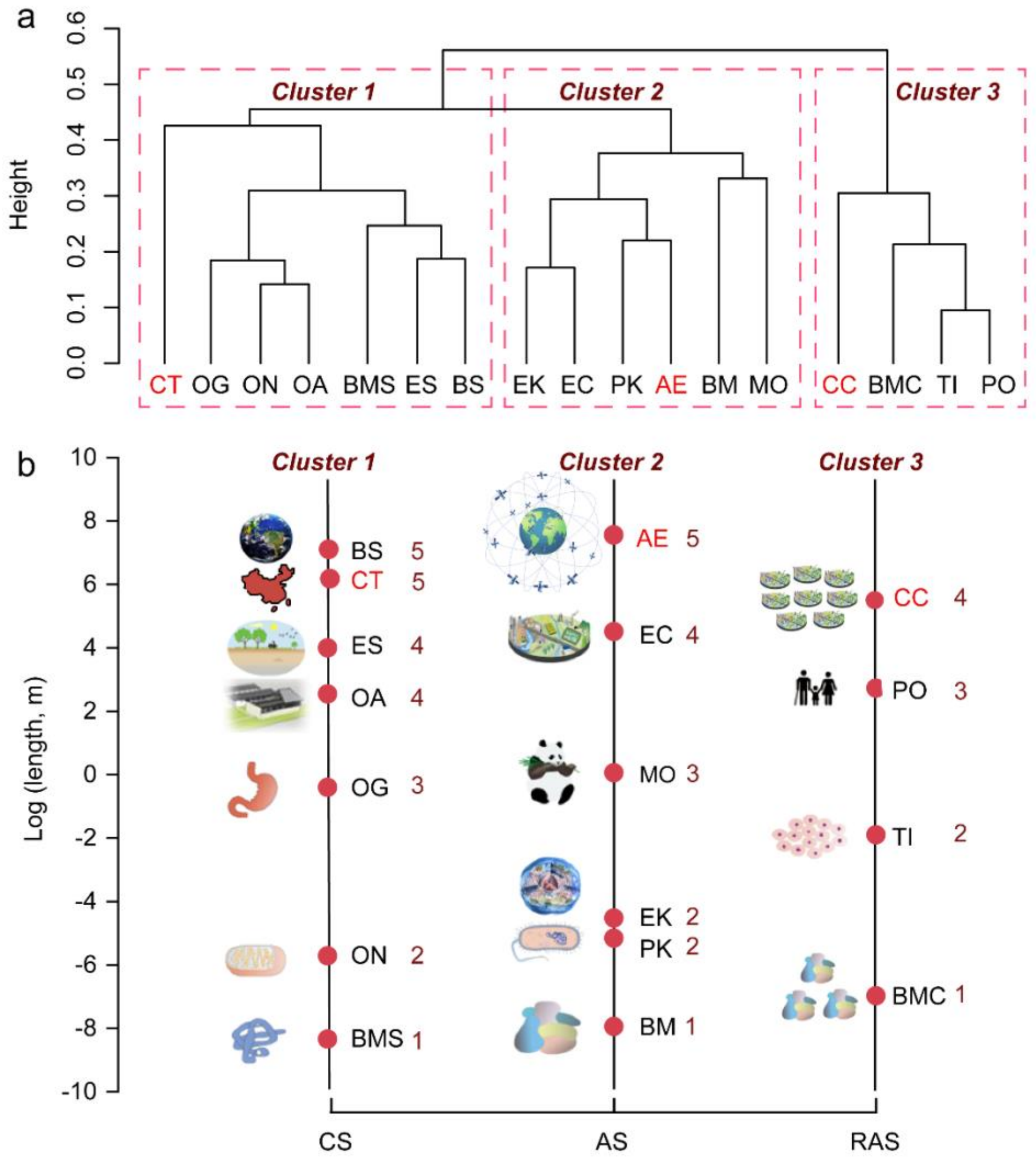

**Fig. S5 Cluster analysis of 17 levels of living systems based on 16 structural, process and functional traits.** (a) Clustering dendrogram, the dashed line marks the cut height of 0.6, which yields three clusters: cluster 1, cluster 2, and cluster 3. (b) Levels arranged by group and plotted against $\log_{10}$ length (m), dark red numbers to the right of system names denote the hierarchical rank within each group. Abbreviations of the known levels see Fig. 1, and CT, country, AE, anthro-Earth, CC, city cluster.

## Supplementary tables

**Table S1 An evaluation of organaras in a eukarcity**

| Classification | Name | Number |
|---|---|---|
| Primary producer | Greenhouse | 3,546 |
| | Vertical farm | 24 |
| Secondary producer | Dairy farm | 490 |
| | Thermal power plant | 15 |
| Decomposer | Sewage treatment plant | 62 |
| | Constructed wetland | 45 |
| Circulator-transporter | Shopping Mall | 490 |
| | Courier station | 88 |
| | Gas station | 75 |
| Life maintainer | Swimming pool | 22 |
| | Gym | 7 |
| | Convenience store | 7 |
| | Restaurant | 437 |
| | Agritourism ecosystem | 666 |
| Regulator | Bank | 131 |
| | Data center | 65 |

**TableS2 Traits of the 14 hierarchical living systems and Ontology ID**

| | Trait | Ontology ID | Term in Ontology |
|---|---|---|---|
| Continuous | | | |
| 1 | Morphology diversity | PATO:0001353 | Structural diversity |
| 2 | Number of components of a system | PATO:0001555 | Number |
| 3 | Metabolism rate in response to system size | SBO:0000545 (or PATO:0000161) | Functional parameter |
| Qualitative | | | |
| 1 | With outer membrane | PATO:0002488 (cross-referenced with GO:0016020) | Composed of |
| 2 | Type of shape | PATO:0000052 | Shape |
| 3 | With hotspots in system | PATO:0000060 | Spatial pattern |
| 4 | Regeneration capacity | PATO:0002123 | Regenerative capacity |
| 5 | Fixed number of components in a system | PATO:0001662 | Constant number |
| 6 | With control center | PATO:0001632 | Centrality |
| 7 | Be controlled by a center | PATO:0001236 (PCO:0000031) | Controlled |
| 8 | Biochemical reaction within a system | PATO:0001236 (cross-referenced with GO:0008152) | Biological process quality |
| 9 | Steady state regulation | PATO:0001546 | Homeostatic |
| 10 | Self-perpetuating unit | PATO:0001433 | Viability |
| 11 | Degree of organization | PATO:0002487 | Structural complexity |
| 12 | Exchange material and energy | PATO:0001534 | Permeability |
| 13 | Genetic information | GENO:0000536 (or PATO:0001321) | Genotype / Hereditary quality |

**Table S3 Ontologies used to define and score traits at each of the 14 known levels of living systems**

| Level | Ontology source | URL |
|---|---|---|
| Biomolecular subunit | CHEBI | https://www.ebi.ac.uk/ols4/ontologies/chebi |
| Biomacromolecule | CHEBI | https://www.ebi.ac.uk/ols4/ontologies/chebi |
| Biomolecular complex | GO | https://www.ebi.ac.uk/ols4/ontologies/go |
| Prokaryote cell | GO | https://www.ebi.ac.uk/ols4/ontologies/go |
| | CL | https://www.ebi.ac.uk/ols4/ontologies/cl |
| | OMP | https://www.ebi.ac.uk/ols4/ontologies/omp |
| Organelle | GO | https://www.ebi.ac.uk/ols4/ontologies/go |
| Eukaryotic cell | GO | https://www.ebi.ac.uk/ols4/ontologies/go |
| | OMP | https://www.ebi.ac.uk/ols4/ontologies/omp |
| | CL | https://www.ebi.ac.uk/ols4/ontologies/cl |
| Tissue | UBERON | https://www.ebi.ac.uk/ols4/ontologies/uberon |
| | FMA | https://www.ebi.ac.uk/ols4/ontologies/fma |
| | PO | https://www.ebi.ac.uk/ols4/ontologies/po |
| Organ | UBERON | https://www.ebi.ac.uk/ols4/ontologies/uberon |
| | PO | https://www.ebi.ac.uk/ols4/ontologies/po |
| Multicellular ;anisms | NCBITaxon | https://www.ebi.ac.uk/ols4/ontologies/ncbitaxon |
| | GO | https://www.ebi.ac.uk/ols4/ontologies/go |
| | PO | https://www.ebi.ac.uk/ols4/ontologies/po |
| Population | PCO | https://github.com/PopulationAndCommunityOntology/pco |
| Ecosystem | ENVO | https://www.ebi.ac.uk/ols4/ontologies/envo |

**Table S4 Scoring criteria for traits of living systems used in cluster analysis**

| Trait | Scale/Scoring | Notes | Trait | Scale/Scoring | Notes |
|---|---|---|---|---|---|
| With outer membrane (layer) | 0 — absent<br>1 — partially present<br>2 — present in most<br>3 — always present | Presence and extent of an outer membrane or boundary layer enclosing the system. | Be controlled by a center | 0 — not<br>1 — Yes | Whether the system is governed by a central control unit. |
| Type of shape | 1 — non-spherical<br>2 — near-spherical<br>3 — spherical | Overall morphological shape of the system unit (e.g., globular vs. irregular). | Metabolism rate in response to system size ($\beta$) | | Metabolic scaling exponent relating metabolic rate to system size. |
| Morphology diversity | $\log_{10}$(richness) | Log-transformed count of distinct morphological forms within the system. | Biochemical reaction within a system | 0 — absent<br>1 — present | Occurrence of internal biochemical reactions within the system boundary. |
| With hotspots in system | 0 — none<br>1 — in a minority<br>2 — in most<br>3 — in all | Regions of concentrated energy and material activity (e.g., organelles in eukaryotic cells). | Steady state regulation | 0 — absent<br>1 — present | Capacity to maintain a stable internal state (homeostasis). |
| Regeneration capacity | 0 — none<br>1 — weak<br>2 — strong | Ability to regenerate or reproduce clonally (e.g., tail regeneration in lizards). | Self-perpetuating unit | 0 — not<br>1 — Yes | Capacity for self-reproduction, self-maintenance, self-determination, and autonomous survival. |

| Trait | Scale/Scoring | Notes | Trait | Scale/Scoring | Notes |
|---|---|---|---|---|---|
| Fixed number of components in a system | 0 — variable number<br>1 — fixed number | E.g., sensory organs in organisms are fixed in number; mitochondria per cell are variable (hundreds to thousands). | Degree of organization | 1 — lowest<br>2 — low<br>3 — high<br>4 — highest | Level of internal coordination and structural integration (e.g., organism = highly organized; population = loosely organized). |
| Number of components of a system | $\log_{10}$ (count) | Log-transformed total count of component units in the system. | Exchange material and energy | 0 — not<br>1 — Yes | Whether the system exchanges material and/or energy with its environment. |
| With control center | 0 — absent<br>1 — in a minority<br>2 — always present | Presence of a dedicated regulatory or coordination center within the system. | Genetic information | 0 — absent<br>1 — partially present<br>2 — always present | Presence of heritable information encoding system properties. Cultural information in cities is considered a novel form of genetic information. |

**Table S5 Qualitative trait scores of the 14 hierarchical living systems used for cluster analysis**

| Living system | With outer membrane (layer) | Type of shape | With hotspots in system | Regeneration capacity | Fixed number of components in a system | With control center |
|---|---|---|---|---|---|---|
| BMS | 0 | 2 | 0 | 0 | 0 | 0 |
| BM | 0 | 3 | 0 | 0 | 1 | 0 |
| BMC | 0 | 2 | 0 | 2 | 0 | 0 |
| PK | 3 | 3 | 0 | 1 | 0 | 1 |
| ON | 2 | 2 | 1 | 1 | 1 | 1 |
| EK | 3 | 3 | 3 | 0 | 0 | 2 |
| TI | 0 | 1 | 0 | 2 | 0 | 0 |
| OG | 2 | 2 | 1 | 1 | 1 | 1 |
| MO | 3 | 1 | 2 | 1 | 1 | 2 |
| PO | 0 | 1 | 0 | 2 | 0 | 0 |
| ES | 0 | 2 | 0 | 2 | 0 | 0 |
| OA | 3 | 2 | 1 | 1 | 1 | 1 |
| EC | 0 | 3 | 3 | 2 | 0 | 2 |
| BS | 2 | 3 | 0 | 2 | 0 | 0 |
| | 0-No; 1-in some types; 2-in large types; 3-in all types | 1-not sphere; 2-near sphere; 3-sphere | 0-No; 1-in some types; 2-in large types; 3-in all types | 0-no; 1-weak; 2-strong | 0-not fixed; 1-fixed | 0-No; 1-in some types; 2-in all types |

**Table S5 (continued) Qualitative trait scores of the 14 hierarchical living systems used for cluster analysis**

| Living system | Be controlled by a center | With biochemical reaction within a system | Steady state regulation | Self-perpetuating unit | Exchange material and energy | Genetic information |
|---|---|---|---|---|---|---|
| BMS | 1 | 1 | 1 | 0 | 1 | 0 |
| BM | 0 | 1 | 1 | 1 | 1 | 0 |
| BMC | 0 | 0 | 0 | 1 | 0 | 0 |
| PK | 0 | 1 | 1 | 1 | 1 | 2 |
| ON | 1 | 1 | 1 | 0 | 1 | 1 |
| EK | 0 | 1 | 1 | 1 | 1 | 2 |
| TI | 0 | 0 | 0 | 0 | 0 | 0 |
| OG | 1 | 1 | 1 | 0 | 1 | 0 |
| MO | 0 | 1 | 1 | 1 | 1 | 0 |
| PO | 0 | 0 | 0 | 0 | 0 | 2 |
| ES | 0 | 1 | 1 | 0 | 1 | 0 |
| OA | 1 | 1 | 1 | 0 | 1 | 1 |
| EC | 0 | 1 | 1 | 1 | 1 | 2 |
| BS | 0 | 1 | 1 | 0 | 1 | 1 |
| | 0-no; 1-yes | 0-no; 1-yes | 0-no; 1-yes | 0-no; 1-yes | 0-little; 1-frequent | 0-No; 1-in some types; 2-in all types |

**Table S6 Quantitative traits of scale of the 14 known levels of living systems**

| Living system | Log length (m) | Distance (Log length, m) | Log mass (g) | Distance (Log mass, g) | Log volume ($m^3$) | Distance (Log volume, $m^3$) | Log time (s) | Distance (Log time, s) |
|---|---|---|---|---|---|---|---|---|
| Biosphere | 7.41 | | 18.34 | | 19.01 | | 11.50 | |
| Eukarcity | 4.56 | 2.84 | 15.78 | 2.56 | 13.45 | 5.56 | 9.00 | 2.50 |
| Organara | 2.40 | 2.17 | 10.34 | 5.44 | 5.48 | 7.97 | 9.00 | 0.00 |
| Ecosystem | 3.00 | | 13.55 | | 8.62 | | 8.00 | |
| Population | 2.70 | 0.30 | 6.89 | 6.66 | 7.29 | 1.32 | 7.00 | 1.00 |
| Multicellular organism | 0.00 | 2.70 | 3.60 | 3.29 | -2.41 | 9.70 | 6.00 | 1.00 |
| Organ | -0.91 | 0.91 | 2.66 | 0.94 | -3.00 | 0.59 | 5.50 | 0.50 |
| Tissue | -1.93 | 1.02 | -0.06 | 2.72 | -6.07 | 3.07 | 5.00 | 0.50 |
| Eukaryote cell | -4.60 | 2.67 | -5.11 | 5.05 | -14.12 | 8.05 | 4.00 | 1.00 |
| Organelle | -5.74 | 1.14 | -12.29 | 7.18 | -17.76 | 3.64 | 3.50 | 0.50 |
| Prokaryote cell | -5.30 | | -11.78 | | -16.18 | | 3.00 | |
| Biomacromolecular complex | -7.73 | 1.99 | -17.70 | 5.42 | -23.48 | 5.73 | -0.50 | 4.00 |
| Biomacromolecule | -8.09 | 0.35 | -18.77 | 1.06 | -24.55 | 1.06 | -3.00 | 2.50 |
| Biomacromolecular subunit | -8.21 | 0.12 | -19.45 | 0.68 | -24.91 | 0.36 | -4.00 | 1.00 |
| **Mean** | | 1.47 | | 3.73 | | 4.28 | | 1.32 |

**Table S7 Qualitative trait of the living systems in Group Component living system**

| | Traits | BMS | ON | OG | ES | OA | BS |
|---|---|---|---|---|---|---|---|
| Structure | Outer membrane | No | Most have | Most have | No | Most have | Most have |
| | Shape | Near sphere | Near sphere | Near sphere | Near sphere | Near sphere | Sphere |
| | Number of components | Varied | Fixed | Fixed | Varied | Fixed | Varied |
| | Hotspot | No | Some types have | A few types | No | Some types have | No |
| | Center | No | A few types | A few types | No | Some types have | No |
| | Genetic information | No | A few types | No | No | Some types have | In some types |
| Main process | Regeneration capacity | No | Weak | Weak | Strong | Weak | Strong |
| | Controlled by a center | Yes | Yes | Yes | No | Yes | No |
| | With biochemical reaction within a system | With | With | With | With | With | With |
| | Steady state regulation | Have | Have | Have | Have | Have | Have |
| | Self-perpetuating unit | Not | Not | Not | Not | Not | Not |
| External relation-ship | Exchange material and energy | Frequent | Frequent | Frequent | Frequent | Freque nt | Freque nt |

Abbreviations see Fig. 1. Criteria see Tables S2 and S4. The judgment was done using the criteria see Supplementary Text S3 for detail.

**Table S8 Qualitative traits of the living systems in Group Autonomous living system**

| | Traits | BM | PK | EK | MO | EC |
|---|---|---|---|---|---|---|
| Structure | Outer membrane | No | Have | Have | Have | No |
| | Shape | Sphere | Sphere | Sphere | Not sphere | Sphere |
| | Number of components | Fixed | Varied | Varied | Fixed | Varied |
| | Hotspot | No | No | Have | In most types | Have |
| | Center | No | A few types | With | With | With |
| | Genetic information | No | Have | Have | No | Have |
| Main process | Regeneration capacity | No | Weak | No | Weak | Strong |
| | Be controlled by a center | No | No | No | No | No |
| | With biochemical reaction within a system | With | With | With | With | With |
| | Steady state regulation | Have | Have | Have | Have | Have |
| | Self-perpetuating unit | Yes | Yes | Yes | Yes | Yes |
| External relation-ship | Exchange material and energy | Frequent | Frequent | Frequent | Frequent | Frequent |

Abbreviations see Fig. 1. Criteria see Tables S2 and S4. The judgment was done using the criteria see Supplementary Text S3 for detail.

**Table S9 Qualitative traits of the living systems in Group Repeat-Autonomous living system**

| | Traits | BMC | TI | PO |
|---|---|---|---|---|
| Structure | Outer membrane | No | No | No |
| | Shape | Near sphere | Not sphere | Not sphere |
| | Number of component | Varied | Varied | Varied |
| | Hotspot | No | No | No |
| | Center | No | No | No |
| | Genetic information | No | No | Yes |
| Main process | Regeneration capacity | Strong | Strong | Strong |
| | Be controlled by a center | No | No | No |
| | With biochemical reaction within a system | No | No | No |
| | Steady state regulation | No | No | No |
| | Self-perpetuating unit | No | No | No |
| External relation-ship | Exchange material and energy | Little | Little | Little |

Abbreviations see Fig. 1. Criteria see Tables S2 and S4. The judgment was done using the criteria see Supplementary Text S3 for details.

**Supplementary references**